\documentclass[twocolumn, english, numberedappendix]{emulateapj}

\usepackage[T1]{fontenc}
\usepackage[latin1]{inputenc}
\usepackage{graphicx}
\usepackage{amssymb}
\makeatletter
\usepackage{times}
\usepackage{amsmath}

\newcommand{\Gyr}{{\,\rm Gyr}}
\newcommand{\Myr}{{\,\rm Myr}}

\newcommand{\pc}{\,\mathrm{pc}}
\newcommand{\kpc}{\,\mathrm{kpc}}

\newcommand{\Mbh}{M_{\bullet}}
 \newcommand{\rs}{R_{\star}}

\newcommand{\Mo}{M_{\odot}}

\shorttitle{SECULAR STELLAR DYNAMICS NEAR AN MBH}
\shortauthors{MADIGAN, HOPMAN AND LEVIN}

\usepackage{babel}
\makeatother
\begin{document}
\bibliographystyle{apj} 

\title{Secular Stellar Dynamics near a Massive Black Hole}

\author{Ann-Marie Madigan$^1$, Clovis Hopman$^1$ and Yuri Levin$^{1,2}$}

\begin{abstract}
The angular momentum evolution of stars close to massive black holes (MBHs) is driven by secular torques. In contrast to two-body relaxation, where interactions between stars are incoherent, the resulting resonant relaxation (RR) process is characterized by coherence times of hundreds of orbital periods. In this paper, we show that all the statistical properties of RR can be reproduced in an autoregressive moving average (ARMA) model. We use the ARMA model, calibrated with extensive $N$-body simulations, to analyze the long-term evolution of stellar systems around MBHs with Monte Carlo simulations. 

We show that for a single-mass system in steady-state, a depression is carved out near an MBH as a result of tidal disruptions. Using Galactic center parameters, the extent of the depression is about 0.1 pc, of similar order to but less than the size of the observed ``hole'' in the distribution of bright late-type stars. We also find that the velocity vectors of stars around an MBH are locally not isotropic. In a second application, we evolve the highly eccentric orbits that result from the tidal disruption of binary stars, which are considered to be plausible precursors of the ``S-stars'' in the Galactic center. We find that RR predicts more highly eccentric ($e > 0.9$) S-star orbits than have been observed to date.\\
\end{abstract}

\keywords{black hole physics -- celestial mechanics -- Galaxy: center -- stars: kinematics and dynamics}

\affiliation{$^1$Leiden Observatory, Leiden University, P.O. Box 9513, NL-2300 RA Leiden, The Netherlands \\$^2$School of Physics, Monash University, Clayton, Victoria 3800, Australia}

\email{madigan@strw.leidenuniv.nl}

\section{Introduction}\label{s:intro}

The gravitational potential near a massive black hole (MBH) is approximately equal to that of a Newtonian point particle. As a consequence, the orbits of stars are nearly Keplerian, and it is useful, both as a mental picture and as a computational device, to average the mass of the stars over their orbits and consider secular interactions between these ellipses, rather than interactions between point particles. These stellar ellipses precess on timescales of many orbits, due to deviations from the Newtonian point particle approximation: there is typically an extended cluster of stars around the MBH, and there is precession due to general-relativistic (GR) effects. Nevertheless, for timescales less than a precession time, torques between the orbital ellipses are coherent.

It was first shown by \citet{Rau96} that such sustained coherent torques lead to much more rapid stochastic evolution of the angular momenta of the stars than normal relaxation dynamics. They called this process resonant relaxation (RR).
RR is potentially important for a number of astrophysical phenomena. \citet{Rau98} showed that it  increases the tidal disruption rate, although in their calculations 
the effect was not very large since most tidally disrupted stars originated at distances that were too large for RR to be effective (in Section \ref{ss:GC_hole} we will revisit this argument). By contrast with tidally disrupted main-sequence stars, inspiraling compact objects originate at distances much closer to the MBH \citep{Hop05}. The effect of RR on the rate of compact objects spiralling into MBHs to become gravitational wave sources is therefore much larger \citep{Hop06a}. RR also plays a role in several proposed mechanisms for the origin of the ``S-stars'', a cluster of B-stars with randomized orbits in the Galactic center \citep[GC; e.g.][see Section \ref{ss:sstars} of this paper]{Hop06a, Lev07, Per07}.

There have been several numerical studies of the RR process itself, which have verified the overall analytic predictions. However, since stellar orbits in $N$-body simulations \citep[e.g.][]{Rau96, Rau98, Aar07, Har08, Eil09, Per09b, Per10} need to be integrated for many precession times, the simulations are computationally demanding. All inquiries have thus far have been limited in integration time and/or particle number. In order to speed up the computation, several papers have made use of the picture described above, where the gravitational interaction between massive wires are considered \citep[Gauss's method, see e.g.][]{Rau96, Gur07, Tou09}. 

It is common, when possible, to use the the Fokker-Planck formalism to carry out long-term simulations of the stellar distribution in galactic nuclei \citep{Bah76, Bah77, Lig77, Mur91, Hop06a,Hop06b}. The current formalism however is not directly applicable to the case when RR plays an
important role. At the heart of all current Fokker-Plank approaches is the assumption of a random-walk diffusion of angular momentum, whereas RR is a more complex relaxation mechanism based on persistent autocorrelations. 

In this work, we will show that the {\it auto-regressive moving average} (ARMA) model provides a faithful representation of all statistical properties of RR.  This model, calibrated with special-purpose $N$-body simulations, allows us to carry out a  study of the long-term effects of RR on the stellar cusp, thus far out of reach. 

The plan of the paper is as follows. In Section \ref{s:nbody_sims}, we present an extensive suite of special purpose $N$-body simulations, which exploit the near-Keplerian nature of stellar orbits and concentrate on the stochastic orbital evolution of several test stars. 
We use these simulations to statistically examine the properties of RR for many secular timescales. In Section \ref{s:stat}, we introduce the ARMA model for the data analysis of the $N$-body simulations. This description accommodates the random, non-secular (non-resonant) effects on very short timescales, the persistent autocorrelations for intermediate timescales less than a precession time, and the random walk behavior for very long times (the RR regime). In Section \ref{ss:theory} we extend the ARMA model, using physical arguments, to a parameter space that is larger than that of the simulations. In Section \ref{s:N} we calibrate the parameters using the results of the $N$-body simulations. Once the free parameters of the ARMA model are determined, we then use them in Section \ref{s:MC}  as an input to Monte Carlo (MC) simulations which study the distribution of stars near MBHs. In Section \ref{s:Results} we show that RR plays a major role in the global structure of the stellar distribution near MBHs, and in particular for the young, massive B-type stars near the MBH (the ``S-stars'') in our GC. We summarize in Section \ref{s:disc}.

\section{$N$-body Simulations }\label{s:nbody_sims}

We have developed a special-purpose $N$-body code, designed to accurately integrate stellar orbits in near-Keplerian potentials; see Appendix \ref{s:app:code} for a detailed description. We wrote this code specifically for the detailed study of RR. Such a study requires an integration scheme with an absence of spurious apsidal precession and one which is efficient enough to do many steps per stellar orbit while integrating many orbits to resolve a secular process. The main features of this code are the following:
\begin{enumerate}
\item A separation between test particles and field particles. Field particles move on Kepler orbits, which precess due to their averaged potential and general relativity, and act as the mass that is responsible for dynamical evolution. Test particles are full $N$-body particles and serve as probes of this potential  \citep{Rau96, Rau98}. 
\item The use of a fourth-order mixed-variable symplectic (MVS) integrator \citep[][]{Yosh90, WH91, Kin91, Sah92}. The MVS integrator switches between Cartesian coordinates (in which the perturbations to the orbit are calculated) to one based on Kepler elements (to calculate the Keplerian motion of the particle under the influence of central object). We make use of Kepler's equation \citep[see, e.g.,][]{Danby92} for the latter step.
\item Adaptive time stepping and gravitational softening. To resolve the periapsis of eccentric orbits \citep{Rau99} and close encounters between particles, we adapt the time steps of the particles. We use the compact $K_2$ kernel \citep{Deh01} for gravitational softening.
\item Time symmetric algorithm. As MVS integrators generally lose their symplectic properties if used with adaptive time stepping, we enhance the energy conservation by time-symmetrizing the algorithm.
\end{enumerate}
Our code is efficient enough to study the evolution of energy and angular momentum of stars around MBHs for many precession times, for a range of initial eccentricities. 

\subsection{Model of Galactic Nucleus}\label{ss:Model_GC}

We base our Galactic nucleus model on a simplified GC template\footnote{Due to the effect of GR precession, the system is not scale-free and the masses need to be specified.}. It has three main components. (1) An MBH with mass $\Mbh = 4 \times 10^6 \Mo$ which remains at rest in the center of the coordinate system. (2) An embedded cluster of equal-mass field stars $m = 10 ~ \Mo$, distributed isotropically from $0.003 \pc$ to $0.03 \pc$, which follow a power-law density profile $n(r) \propto r^{-\alpha}$. The outer radius is chosen with reference to \citet{Gur07}, who show that stars with semi-major axes larger than the test star's apoapsis distance $r_{\rm apo} = a (1 + e )$ contribute a negligible amount of the net torque on the test star. The field stars move on precessing Kepler orbits, where the precession rate is determined by the smooth potential of the field stars themselves (see Appendix \ref{s:app:prec}) and general relativity. The precession of the field stars is important to account for because for some eccentricities the precession rate of the test star is much lower than that of the ``typical'' field star, such that it is the precession of the latter that leads to decoherence of the system. The field stars do not interact with each other but they do interact with the test stars if the latter are assumed to be massive. In this way the field stars provide the potential of the cluster but are not used as dynamic tracers. (3) A number of test stars, used as probes of the background potential, that are either massless or have the same mass as the field stars, $m = 10 \Mo$. We consider both massless and massive stars in order to study the effects of resonant friction \citep{Rau96}. The test stars have semi-major axes of $a = 0.01 \pc$ and a specified initial eccentricity $e$. 

Following Equation (17) from \citet{Hop06a}, who use the $\Mbh\!\--\!\sigma$ relation \citep{Fer00,Geb00} which correlates the mass of a central black hole with the velocity dispersion of the host galaxy's bulge, we define the radius of influence as 
\begin{equation}
r_h = { G \Mbh \over \sigma^2} = 2.31 \pc \left( {\Mbh \over 4 \times 10^6 \Mo} \right)^{1/2}.
\end{equation}
The number of field stars within radius $r$ is
\begin{equation}\label{e:N_h}
N(< r) = N_h \left( r \over r_h\right) ^{3 - \alpha},
\end{equation}
where we assume that the mass in stars within $r_h$ equals that of the MBH,  $N_h = \Mbh/m = 4 \times 10^5$. We take $\alpha = 7/4$ \citep{Bah76}, the classic result for the distribution of a single-mass population of stars around an MBH, which relies on the assumption that the mechanism through which stars exchange energy and angular momentum is dominated by two-body interactions. Hence the number of field stars within our model's outer radius is $N ( < 0.03 \pc) = 1754$. We summarize the potential for our Galactic nucleus model in Table \ref{t:gn}. 

\begin{table} [h]
\caption{Galactic Nucleus Model}
\begin{center}
\begin{tabular}{cc}
\hline
\hline 
Parameter & Numerical Value \tabularnewline
\hline
$\Mbh$  & $4 \times 10^6 \Mo$	\tabularnewline
$m$ &  $10 ~\Mo$	\tabularnewline
$\alpha$ &  1.75 \tabularnewline
$r_h$ & $2.31 \pc$ \tabularnewline
$r_{\rm min}$ & $0.003 \pc$ \tabularnewline
$r_{\rm max} $& $0.03 \pc$ \tabularnewline
$a_{\rm test star}$ & $0.01 \pc$ \tabularnewline
$N (< r_{\rm max})$ & $1754$ \tabularnewline
\hline
\end{tabular}
\end{center}
\label{t:gn}
\end{table}

We evolve this model of a galactic nucleus for a wide range in eccentricity of the test stars, $e = 0.01, 0.1, 0.2, 0.3, 0.4, 0.6, 0.8, 0.9$ and $0.99$. For each initial eccentricity, we follow the evolution of a total of 80 test stars, both massless and massive, in a galactic nucleus. Typically we use four realizations of the surrounding stellar cluster for each eccentricity, i.e., 20 test stars in each simulation. They have randomly oriented orbits with respect to one another, so that they experience different torques within the same cluster. The simulations are terminated after $6000$ orbital periods (henceforth denoted by $P = 2 \pi \sqrt{a^3/G \Mbh}$) at $a = 0.01 \pc$, roughly equivalent to three precession times for a star of eccentricity $e = 0.6$, deep into the RR regime for all eccentricities; see Section \ref{s:N} for verification. Using our method of analysis, it would not be useful for our simulations to run for longer times as the stars would move significantly away from their initial eccentricities. In addition, the autocorrelation functions (ACFs) of their angular momentum changes drop to zero before this time; see Figure \ref{f:acf}.

\subsection{Illustrative simulations}\label{ss:Sim_Results}

In Section \ref{s:stat}, we present a description that captures all the relevant statistical properties of RR, and in particular has the correct autocorrelations. Here we consider several individual simulations for illustration purposes, in order to highlight some interesting points and motivate our approach.
We define energy $E$ of the test star as 
\begin{equation}\label{e:E_test}
E = {G \Mbh \over 2 a} ,
\end{equation}
and dimensionless angular momentum\footnote{Throughout this paper, we use units in which angular momentum $J$ and torque $\tau$ are normalized by the circular angular momentum for a given semi-major axis, $J_c = \sqrt{G \Mbh a}$. All quantities are expressed per unit mass.} $J$ as
 \begin{equation}\label{e:J_test}
J = \sqrt{1 - e^2} .
\end{equation}

\begin{figure}
	\begin{center}
			\includegraphics[height=85 mm, angle=270]{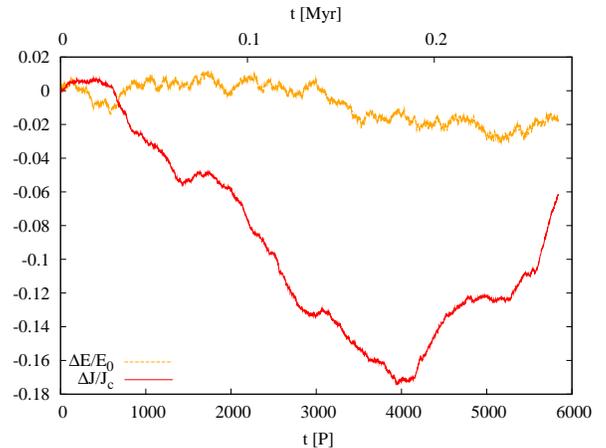}
		\caption{ Evolution of angular momentum $J$ and energy $E$, normalized to the initial values, of a massless test star with semi-major axis $a = 0.01 \pc$ and eccentricity $e = 0.6$. Time is in units of the initial orbital period $P$ (top axis shows time in Myr). The coherence of RR can be seen in the $J$ evolution (precession time $t_{\rm prec}$ is $\sim 2000$ orbits), whereas the $E$ evolution displays a much slower, more erratic, random walk.
		 \label{f:sim_0.6EJ}}
	\end{center}
\end{figure}

Secular torques should affect the angular momentum evolution, but not the energy evolution of the stars (in the picture of interactions between massive ellipses of the introduction, the ellipses are fixed in space for times less than a precession time,  $t < t_{\rm prec}$, such that the potential and therefore the energy is time-independent). It is therefore of interest to compare the evolution of these two quantities. An example is shown in Figure \ref{f:sim_0.6EJ}. As expected, the angular momentum evolution is much faster than energy evolution; furthermore, the evolution of angular momentum is much less erratic, which visualizes the long coherences between the torques.

\begin{figure}
	\begin{center}
			\includegraphics[height=85 mm,angle=270 ]{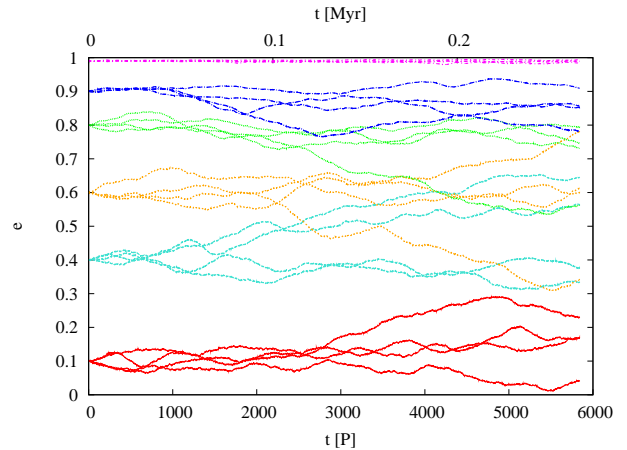}
		\caption{Eccentricity evolution of a sample of massless test stars in a fiducial galactic nucleus model. We show four realizations for five different initial eccentricities. Evolution is rapid for most eccentricities; however, the stars with the highest value $e = 0.99$ have sluggish eccentricity evolution. 
		 \label{f:sim_e_time}}
	\end{center}
\end{figure}

In Figure \ref{f:sim_e_time} we show the eccentricity evolution for a sample population of stars with various initial orbital eccentricities. There is significant eccentricity evolution in most cases, but almost none for $e = 0.99$, the reasons for which we elucidate in Section \ref{s:MC}. 

To quantify the rate at which the energy $E$ and angular momentum $J$ change as a function of eccentricity $e$, we calculate the $E$ and $J$ relaxation timescales. We define these as the timescale for order of unity (circular angular momentum) changes in energy $E$ (angular momentum $J$). We compute the following quantities:
 \begin{equation}\label{e:diffusion_E}
t_E =  {E ^2 \over \left\langle (\Delta E)^2 \right\rangle} \Delta t,
\end{equation}
 \begin{equation}\label{e:diffusion_J}
 t_J =  {J_c ^2 \over \left\langle (\Delta J)^2 \right\rangle} \Delta t,
\end{equation}
where $\Delta E$ and $\Delta J$ are the steps taken in a time $\Delta t$, and we take the mean $\langle \,\rangle$ over 80 test stars in each eccentricity bin. 

\begin{figure}
	\begin{center}
			\includegraphics[height=85 mm,angle=270 ]{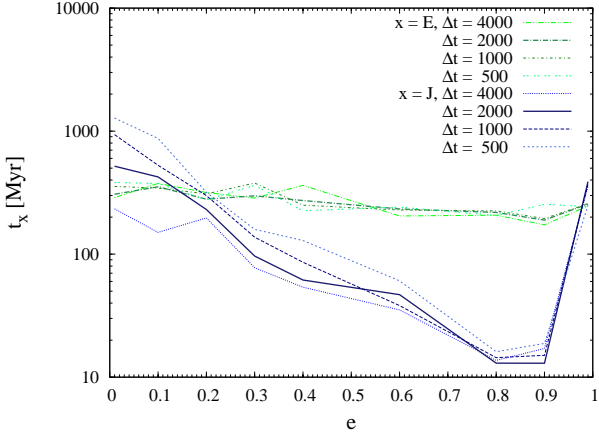}
		\caption{Timescale for order of unity (circular angular momentum) changes in energy $E$ (angular momentum $J$) for massless test stars in our $N$-body simulations, presented on a log scale of time in units of Myr; see Equations (\ref{e:diffusion_E}) and (\ref{e:diffusion_J}). We find a weak trend for lower $E$ relaxation times with increasing eccentricity (from $\sim 400 \Myr$ to $200 \Myr$). $J$ relaxation occurs on very short timescales ($\sim 20 \Myr$) at eccentricities of $0.8 < e < 0.9$. Stellar orbits with low eccentricities have very long $J$ relaxational timescales ($\sim 1 \Gyr$) as the torque on a circular orbit drops to zero. This timescale increases again at very high eccentricities.
		 \label{f:trr_tnr}}
	\end{center}
\end{figure}

We plot the resulting timescales as a function of eccentricity in Figure \ref{f:trr_tnr}, taking several $\Delta t$ values to get order-of-magnitude estimates for these timescales. We find no significant difference between the results for the massless and massive test stars. We note that $t_E$ is only weakly dependent on $e$, whereas $t_J$ is of the same order of magnitude as $t_E$ for $e\to0$ and $e\to1$, but much smaller for intermediate eccentricities. It is in the latter regime that secular torques are dominating the evolution. In the following sections, we will give a detailed model for the evolution of angular momentum. Since the focus of this paper is mainly on secular dynamics, we do not further consider energy diffusion here, but in Appendix \ref{s:app:energy} we discuss the timescale for cusp formation due to energy evolution, and compare our results to other results in the literature.

\section{Statistical description of resonant relaxation}\label{s:stat}

As a result of the autocorrelations in the changes of the angular momentum of a star, as exhibited in Figure \ref{f:sim_0.6EJ}, RR cannot be modeled as a random walk for all times. This is also clear from physical arguments. \citet{Rau96} and several other papers take the approach of considering two regimes of evolution: for times $ t \ll t_{\rm prec}$, evolution is approximately linear as the torques continue to point in the same direction. For times $t \gg t_{\rm prec}$, the torque autocorrelations vanish and it becomes possible to model the system as a random walk. Here we introduce a new approach, which unifies both regimes in a single description. This description is also useful as a way of quantifying the statistical properties of RR. 

\begin{figure}
	\begin{center}
			\includegraphics[height=85 mm,angle=270 ]{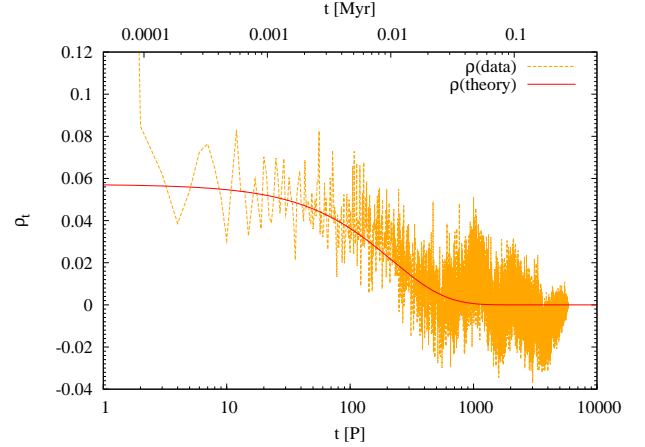}
		\caption{Autocorrelation function for the differences $\Delta J$ in one particular simulation (noisy orange line) as a function of time in units of the initial orbital period $P$. The test star is massless and has initial eccentricity $e = 0.6$. The autocorrelation function is significantly larger than zero (if much less than one) for several hundred orbits. The smooth red line gives the theoretical ACF for an ARMA(1, 1) model with $\phi_1=0.995$, $\theta_1=-0.976$. \label{f:acf}}
	\end{center}
\end{figure}

We study the evolution of the angular momenta of the test stars using a time series of angular momenta at a regular spacing of one period. This choice is arbitrary, and below we show how our results generalize to any choice of time steps. The normalized ACF can be written as

\begin{equation}
\rho_t = {\langle (\Delta J_{s+t} - \langle \Delta J \rangle)(\Delta J_{s}- \langle \Delta J \rangle)\rangle\over\langle (\Delta J_{t} - \langle \Delta J \rangle)^2\rangle},
\end{equation}
where $\Delta J_{s}$ is the change in angular momentum between subsequent measurements at time $s$ and it is assumed that the time series is stationary. A typical ACF from our simulations is shown in Figure \ref{f:acf}. The fundamental feature of this ACF is that it is significantly larger than zero for hundreds of orbits, but with values much less than unity.

Our goal in this section is to define a process that generates a time series that has the same ACF as that of the changes in angular momentum of a star. Such a time series has all the statistical properties that are relevant for the angular momentum evolution of the star, such as the same Fourier spectrum. We have two motivations for doing this. First, defining such a process implies that once the parameters of the model are calibrated by use of the $N$-body simulations, RR is fully defined. Second, a generated time series with the same statistical properties as the time series generated in the physical process can be used in MC simulations to solve for the long term evolution of RR.

\subsection{The autoregressive moving average model ARMA(1, 1)}

We now introduce the ARMA model, which is often used in econometrics \citep[see, e.g.,][]{Heij04}. We show that this model can generate time series that reproduce the ACF and variance of $\Delta J_t$. The model is ad hoc in that it does not have a direct physical foundation. However, we will find physical interpretations for the free parameters of the model in Section \ref{ss:theory}. We use a form of the model, ARMA(1,1), with one autoregressive parameter, $\phi_1$, and one moving average parameter, $\theta_1$. The model can then be written as

\begin{equation}\label{e:DJ1}
\Delta_1 J_t = \phi_1\Delta_1 J_{t-1} + \theta_1\epsilon_{t-1}^{(1)} + \epsilon_t^{(1)}.
\end{equation}
The label ``1'' in this equation refers to the fact that the data have a regular spacing of one period. In this equation, $\phi_1$ and $\theta_1$ are free parameters, and the random variable $\epsilon$ is drawn from a normal distribution with

\begin{equation}\label{e:iid}
\langle\epsilon^{(1)}\rangle=0;\quad \langle\epsilon_t^{(1)}\epsilon_s^{(1)}\rangle=\sigma_1^2\delta_{ts}
\end{equation}
where $\sigma_1$ is a third free parameter and $\delta_{ts}$ is the Kronecker delta. For such a model, the variance of the angular momentum step is
\begin{equation}\label{e:sigvar}
\langle \Delta J_t^2\rangle = {1 + \theta_1^2 + 2\theta_1\phi_1\over 1-\phi_1^2}\sigma_1^2,
\end{equation}
and the ACF can be derived analytically (see Appendix \ref{s:app:eqns} for details):
\begin{equation}\label{e:acf}
\rho_t = \phi_1^t\left[1 + {\theta_1/\phi_1\over 1 + (\phi_1+\theta_1)^2/(1-\phi_1^2)}\right]\quad\quad(t>0) .
\end{equation}
From this expression, we see that the decay time of the ACF is determined by the parameter $\phi_1$, and $\theta_1$ captures the magnitude of the ACF. As an example, we plot the ACF in Figure \ref{f:acf} with parameters typical for RR. We emphasize the need for both an autoregressive parameter and a moving average parameter to reproduce the slow decay of the ACF. Neither parameters can replicate the features on their own.

A useful reformulation of the ARMA(1, 1) model, which can be found using recursion, is 
\begin{equation}
\Delta J_t = \sum_{k=0}^{\infty}\psi_k \epsilon_{t-k},
\end{equation}
where
\begin{equation}
\psi_k = (\phi_1+\theta_1)\phi_1^{k-1}\quad (k>0); \quad \psi_0 = 1.
\end{equation}

\begin{figure}
	\begin{center}
			\includegraphics[height=85 mm,angle=270 ]{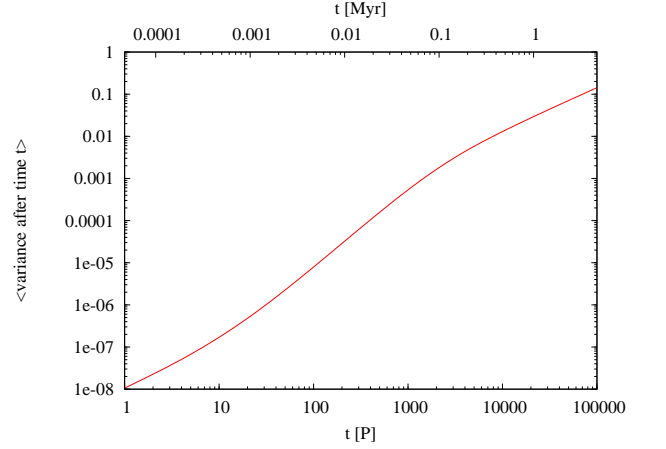}
		\caption{Variance in angular momentum $J$ after time $t$ ($V(t)$; from Equation (\ref{e:DX})) as a function of time in units of the initial orbital period $P$, with $\phi_1=0.999$, $\theta_1=-0.988$ and $\sigma_1=10^{-4}$. Note that there is first a linear part (``NR''), then a quadratic part (``coherent torque''), and then again a linear part (``RR''). \label{f:DX}}
	\end{center}
\end{figure}

Once the model parameters $(\phi_1, \theta_1, \sigma_1)$ are found for a time step of one period, $\delta t=P$, the variance after $N$ steps can be computed. The displacement after $N$ steps is given by
\begin{equation}
\sum_{n=1}^{N}\Delta J_n = \sum_{n=1}^{N}\sum_{k=0}^{\infty}\psi_k\epsilon_{n-k},
\end{equation}
where $(n - k ) \geq 0$. We denote the variance of the displacement after $N$ steps by
\begin{equation}\label{e:VN}
V(N) \equiv \left\langle\left(\sum_{n=1}^{N}\Delta J_n\right)^2\right\rangle,
\end{equation}
such that
\begin{eqnarray}
V(N) &=& \sum_{n=1}^{N}\sum_{k=0}^{\infty}\sum_{m=1}^{N}\sum_{l=0}^{\infty}\psi_k\psi_l\langle\epsilon_{n-k}\epsilon_{m-l}\rangle\nonumber\\
&=& \sigma_1^2\sum_{n=1}^{N}\sum_{k=0}^{\infty}\sum_{m=1}^{N}\sum_{l=0}^{\infty}\psi_k\psi_l\delta_{n-k, m-l},
\end{eqnarray}
which can be decomposed as
\begin{equation}\label{e:DXN}
\begin{split}
V(N)\sigma_1^{-2} = N &+ \sum_{m=2}^{N}\sum_{n=1}^{m-1}\psi_{m-n} + \sum_{k=1}^{N-1}\psi_k(N-k) \\
&+ \sum_{k=1}^{\infty}\sum_{m=1}^{N}\sum_{n=1}^{\min(k+m-1, N)}\psi_k\psi_{k+m-n}.
\end{split}
\end{equation}
This expression can be cast in a form without summations by repeatedly using the properties of the independent normal random variables $\epsilon_t$ in Equation (\ref{e:iid}), resulting after some algebra in
\begin{equation}\label{e:DX}
\begin{split}
V(N)\sigma_1^{-2} = N &+ {\phi_1+\theta_1\over1-\phi_1}\left[2N-1 + {2\phi_1^N - \phi_1 -1\over1-\phi_1} \right] \\
&+ \left({\phi_1+\theta_1\over1-\phi_1}\right)^2\left[N - 2{\phi_1(1-\phi_1^{N})\over1-\phi_1^2}\right].
\end{split}
\end{equation}
Note that in the special case that $N=1$, Equation (\ref{e:sigvar}) is recovered. We plot this expression in Figure \ref{f:DX}.

Three characteristic timescales are clearly discernible in this figure. On a short timescale, the angular momentum diffusion is dominated 
by the usual two-body gravitational scattering (i.e., non-resonant relaxation, hereafter NR). In this regime the angular momentum deviation scales as $\sqrt{t}$. On an intermediate timescale the orbital angular momentum evolves due to the nearly constant orbit-averaged secular torques from the other stellar orbits, and thus the angular momentum deviation scales linearly with time. On a yet longer timescale the secular torques fluctuate due to orbital precession, and the $\sqrt{t}$ scaling is reinstated, albeit with a much higher effective diffusion coefficient. Thus the ARMA-driven evolution is in agreement with what is expected for a combination of RR and NR acting together; see also \cite{Rau96} and \cite{Eil09} for discussion.

For the typical values of $\phi_1$ and $\theta_1$ that we find, these different regimes can be identified in the expression for the variance $V(N)$; Equation (\ref{e:DX}). For very short times, the expression is dominated by the first term, which can thus be associated with NR. The second term and third term are both proportional to $N^2$ for short times (this can be verified by Taylor expansions assuming $N\phi_1\ll1$; the linear terms cancel), and as such represent coherent torques; and they become proportional to $N$ for large time, representing RR. In Figure \ref{f:DX} we plot Equation (\ref{e:DX}) for typical values of the parameters of the model.

\begin{figure}
	\begin{center}
			\includegraphics[height=85 mm,angle=270 ]{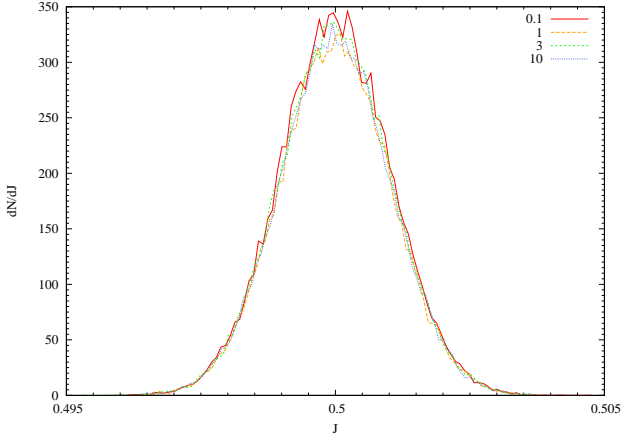}
		\caption{Monte Carlo simulation of RR with the model parameters $\phi_1=0.999$, $\theta_1=-0.988$, and $\sigma_1=10^{-6}$ (chosen for purely illustrative purposes), and the time steps in units of the orbital period indicated in the legend. After $10^4$ orbits, the distribution is very similar for all time steps chosen, confirming that the diffusion in $J$ is independent of the time step. \label{f:ink}}
	\end{center}
\end{figure}

Equation (\ref{e:DX}) is useful for finding the variance after $N$ periods, but most importantly to define a new process with arbitrary time steps. We can now define a new ARMA(1, 1) process with larger time steps, but with the same statistical properties. If the new process makes steps of $N$ orbits each, we write the new model as
\begin{equation}\label{e:DJN}
\begin{split}
\Delta_N J_t &= \phi_N\Delta_N J_{t-N} + \theta_N\epsilon_{t-N}^{(N)} + \epsilon^{(N)}_t ;\\
\langle [\epsilon^{(N)}]^2\rangle &= \sigma_N^2,
\end{split}
\end{equation}
where the parameters $[\phi(\delta t), \theta (\delta t )]$ are related to $(\phi_1, \theta_1)$ by\footnote{For $\phi_1$ we use the fact that the theoretical ACF of an ARMA(1, 1) model decays on a timescale $P/\ln\phi_1$. We have no interpretation for the functional form of $\theta_1$, which was found after some experimenting. We confirm numerically that this form leads to excellent agreement with simulations with arbitrary time; see Figure \ref{f:ink}.}
\begin{equation}\label{e:pt}
\phi(\delta t) = \phi_1^{\delta t/P};\quad\quad \theta(\delta t) = -[-\theta_1]^{\delta t/P},
\end{equation}
and in particular, the ARMA parameters for steps of $N$ orbits are
\begin{equation}\label{e:pN}
\phi_{N} = \phi(NP) = \phi_1^{N};\quad\quad \theta_{N} = \theta(NP) = -[-\theta_1]^{N}.
\end{equation}

Since after $N$ small steps of one period, the variance should be the same as after 1 single step of time $N$ orbits, we can find the new $\sigma$ to be (see Equations (\ref{e:sigvar}), (\ref{e:DX}))
\begin{equation}\label{e:sigmaN}
\sigma_N^2 = {1-\phi_N^2\over 1+\theta_N^2 + 2\phi_N\theta_N} V (N).
\end{equation}

Once the ARMA parameters for steps of one orbital period are found, they can be generalized to a new ARMA model via Equation (\ref{e:DJN}), so that the time step equals an arbitrary number of periods (not necessarily an integer number), and the new parameters are found by the use of Equations (\ref{e:pN}) and (\ref{e:sigmaN}). We will exploit these relations in MC simulations (Section \ref{s:MC}), in which we use adaptive time steps. We have tested the equivalence of the models by considering an ARMA diffusion process over a fixed time interval but with varied time steps. We show an example in Figure \ref{f:ink} where we have taken a delta function for the initial $J$-distribution, and have plotted the resulting final distributions after $10^4$ orbits for several different time steps which vary by two orders of magnitude. We find excellent agreement between these distribution functions which verify the generalization of $(\phi_1, \theta_1, \sigma_1)$ to $(\phi_N, \theta_N, \sigma_N)$.

\section{Interpretation of the ARMA parameters and extension of parameter space}\label{ss:theory}

The ARMA(1, 1) model presented here contains the three free parameters $(\phi_1, \theta_1, \sigma_1)$. In this section, we interpret these parameters in the context of a stylized model of a galactic nucleus. Our purpose in doing so is twofold. First, it relates the ARMA model to the relevant physical processes that play a part in the angular momentum evolution of stars near MBHs, and gives insight into how the parameters depend on the stellar orbits, in particular on their eccentricity. This allows us to define, and discuss, RR in the language of the ARMA model. Second, in Section \ref{s:N} we calibrate the parameters as a function of eccentricity using $N$-body simulations for one specific configuration --- as specified in Section \ref{ss:Model_GC}. In order for the parameters to be also valid for other conditions than those studied, one needs to understand how they vary as a function of the quantities that define the system. The model presented in this section, which is based on physical arguments, allows us to generalize our results to models we have not simulated (models with different stellar masses, semi-major axes, etc.) The arguments presented here do not aspire to give a full theoretical model of RR, but rather parameterize the model in a simplified way that is useful for generalization to other systems. We will use this model in our MC simulations (Section \ref{s:MC}).

\subsection{Non-resonant relaxation (NR): The parameter $\sigma$}

For very short times NR dominates the angular momentum evolution of a star and the variance of $J$ is expected to be of order $P/t_{\rm NR}$ after one period, where

\begin{equation}\label{e:tNR}
t_{\rm NR} = A_{\rm NR} \left({\Mbh\over m}\right)^2{1\over N_<}{1\over \ln\Lambda}P
\end{equation}
is the NR time \citep{Rau96}, defined as the timescale over which a star changes its angular momentum by an order of the circular angular momentum at that radius. In this expression, $N_<$ is the number of stars within the radius equal to the semi-major axis of the star, $\Lambda=\Mbh/m$, and $A_{\rm NR} = 0.26$, chosen to match the value of $t_E$ from the $N$-body simulations; see Figure \ref{f:trr_tnr}. Since this variance should be approximately equal to $\sigma_1^2$, it follows that
\begin{equation}\label{e:sigmath}
\sigma_1 = f_{\sigma}\left({m\over\Mbh}\right)\sqrt{ {N_<\ln\Lambda \over A_{\rm NR}} }.
\end{equation}
Here $f_\sigma$ is a (new) free parameter which, calibrated against our $N$-body simulations, we expect to be of order unity (the deviation from unity will quantify a difference in $t_E$ and NR component of $t_J$). In our $N$-body simulations we find that $\sigma_1$ is an increasing function of $e$; see Figures \ref{f:sigma_2m} and \ref{f:sigma}. This suggests that NR is also an increasing function of $e$ which we attribute to the increase in stellar density at small radii which a star on a highly eccentric orbit passes through at periapsis. To account for this in our theory we fit $f_{\sigma}$ as a function of $e$.

\subsection{Persistence of coherent torques: The parameter $\phi$}

We now define a new timescale over which interactions between stars remain coherent, the {\it coherence time} $t_{\phi}$, which corresponds to the time over which secular torques between orbits remain constant. This is the minimum between the precession time of the test star and the median precession time of the field stars. The latter timescale is of importance as there exists at every radius an orbital eccentricity which has equal, but oppositely directed, Newtonian precession due to the potential of the stellar cluster and GR precession. A star with this particular orbital eccentricity remains almost fixed in space; for our chosen model of a galactic nucleus the eccentricity at which this occurs is $e \sim 0.92$ at $0.01 \pc$. Its coherence time, however, does not tend to infinity, as the surrounding stellar orbits within the cluster continue to spatially randomize.

The timescale for the Newtonian precession (i.e., the time it takes for the orbital periapsis to precess by $2 \pi$ radians) is given by
\begin{equation}\label{e:tcl}
t^{\rm cl}_{\rm prec} = \pi (2 - \alpha) \left( {\Mbh \over N_< m} \right) P(a) f(e, \alpha)^{-1}
\end{equation}
(see Appendix \ref{s:app:prec} for derivation), where $N_<$ is the number of stars within the semi-major axis $a$ of the test star, $P(a)$ its period, and
\begin{equation}\label{e:f(e)}
\begin{split} 
f &(e, \alpha = 7/4)^{-1}  \\
&= \left[ 0.975(1-e)^{-1/2}+0.362(1-e)+0.689 \right].
\end{split} 
\end{equation}
The timescale for precession due to general relativity is given by
\begin{equation}\label{e:tGR}
t^{\rm GR}_{\rm prec} = {1\over3}(1-e^2){ac^2\over G\Mbh}P(a)
\end{equation}
\citep{Ein16}. The combined precession time is
\begin{equation}\label{e:tprec}
t_{\rm prec}(a, e) = \left|{1\over t^{\rm cl}_{\rm prec}(a, e)} - {1\over t^{\rm GR}_{\rm prec}(a, e)}\right|^{-1}.
\end{equation}

As stated, the coherence time for a particular star is the minimum of its precession time and that of the median precession time of the surrounding stars,
\begin{equation}\label{e:tphi}
t_{\phi} = f_{\phi}\min\left[t_{\rm prec}(a,e), t_{\rm prec}(a,\tilde{e})\right] ,
\end{equation}
where $\tilde{e}$ is the median value of the eccentricity of the field stars (so $\tilde{e}=\sqrt{1/2}$ for a thermal distribution) and $f_\phi$ is a second (new) free parameter. 

We now have all the ingredients to define the RR time. This is the timescale over which the angular momentum of a star changes by order of the circular momentum. The angular momentum evolution is coherent over a time of order $t_\phi$, and the change in angular momentum during that time is $\tau t_\phi$, where $\tau$ is the torque exerted on the star's orbit. Normalized to the circular angular momentum, this torque is \citep{Rau96}
\begin{equation}\label{e:tau}
\tau = A_{\tau} {m\over\Mbh}{\sqrt{N_<}\over P} e,
\end{equation}
where the linear dependence on eccentricity was determined by \citet{Gur07}. We adopt their value of $A_{\tau} = 1.57$ which was determined for a different stellar mass profile $m = 1 \Mo, \alpha = 1.4$ where $\Mbh = 3 \times 10^6 \Mo, r_h = 2 \pc$. Assuming that the evolution is random for longer times, this leads to the definition
\begin{equation}\label{e:tRR}
t_{\rm RR}(E, J)=\left({1\over \tau t_\phi}\right)^2 t_\phi.
\end{equation}

The theoretical ACF of an ARMA(1, 1) model in Equation (\ref{e:acf}) shows that it decays on a timescale $P/\ln\phi_1$. Hence we find that $\phi$ scales as $\exp (-{P/ t_{\phi}} )$. However, even if the coherence time is very long, the evolution of the orbit may not be driven by secular dynamics. The secular torques on an orbit are proportional to its eccentricity so if the eccentricity is very small secular effects are weak. As a result, the evolution is dominated by two-body interactions, which have a vanishingly short coherence time. Within our formalism, this effect can be accounted for by multiplying $t_{\phi}$ by a function $S$, 
\begin{equation}\label{e:phith}
\phi_1 = \exp\left(-{P\over S t_{\phi}}\right)
\end{equation}
with
\begin{eqnarray}\label{e:S}
S &=& {1 \over 1 + \exp(k ~ [e - e_{\rm crit}])}
\end{eqnarray}
which interpolates between the secular and two-body regime. We remind the reader that the subscript ``1'' is used to emphasize that this is the value of $\phi$ for a time step of one orbital period. The exponential transition between NR and RR regimes is somewhat arbitrarily chosen but fits the data reasonably well. $k > 0$ is a new parameter which determines the steepness of the transition (a third free parameter), and $e_{\rm crit}$ is defined as the critical eccentricity at which the NR and RR times are equal (Equations (\ref{e:tNR}), (\ref{e:tRR})):
\begin{equation}\label{e:e_crit}
e_{\rm crit}(a,e) =\sqrt{ {\ln \Lambda \over A_{\rm NR} A_{\tau}^2} }\left({P \over t_{\phi}}\right)^{1/2}.
\end{equation}

\begin{figure}
\begin{center}
			\includegraphics[height=85 mm,angle=270 ]{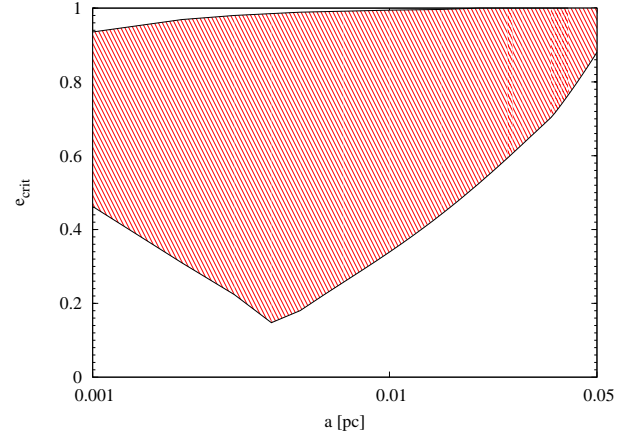}
		\caption{Critical eccentricity $e_{\rm crit}$ for which NR and RR mechanisms work on comparable timescales, as a function of semi-major axis $a$ from the MBH in our Galactic nucleus model. The shaded region in this plot indicates the eccentricities for which RR is the dominant mechanism. The lower values of $e_{\rm crit}$ decrease with distance to the MBH as RR becomes more effective ($\tau$ and $t_{\phi}$ increase with decreasing $a$). At $a \sim 0.004 \pc$ this value rises again in our model as $t_{\rm NR}$ decreases rapidly as $r^{-1/4}$. The higher values of $e_{\rm crit}$ are due to GR effects which decrease $t_{\phi}$ toward the MBH and substantially reduce the effectiveness of RR. The lower values of $e_{\rm crit}$ are due to the fact that $\tau \propto e$.
		\label{f:ecrit}}
		\end{center}
\end{figure}

Inserting relevant values into the above equation and solving numerically using a Newton-Raphson algorithm, we calculate $e_{\rm crit} \sim 0.3$ at $a = 0.01 \pc$ which is in good agreement with the $N$-body simulations. We plot $e_{\rm crit}$ as a function of $a$ in Figure \ref{f:ecrit}. RR becomes more effective at lower $a$ as the mass precession time increases, and the value of $e_{\rm crit}$ decreases. At the lowest $a$ there are two roots to Equation (\ref{e:e_crit}); this is due to GR effects which compete with mass precession and decrease the coherence time to such an extent as to nullify the effects of RR at high eccentricities. These results are derived for a simplified galactic nucleus model and will change for different assumptions on the mass distribution; nevertheless we find the concept of $e_{\rm crit}$ useful for understanding the relative importance of RR and NR at different radii.

\subsection{Magnitude of resonant relaxation steps: The parameter $\theta$}
For times less than $t_\phi$, there are coherent torques between stellar orbits; see Equation (\ref{e:tau}). The expected variance after a time $t_{\phi}$ is then
\begin{equation}\label{e:DJphi}
\langle \Delta J_\phi^2 \rangle = A_{\tau}^2  \left({m\over\Mbh}\right)^2N_<\left({t_{\phi}\over P}\right)^2e^2.
\end{equation}
Alternatively, we can use the ACF (see Appendix \ref{s:app:eqns} for derivation) to find
\begin{equation}\label{e:DJphiacf}
\langle \Delta J_\phi^2 \rangle = \sigma_1^2 \left( {t_{\phi} \over P} \right)^2 \frac{(\theta_1 + \phi_1) ( \theta_1 + 1/\phi_1)}{1 - \phi_1^2} .
\end{equation}
Equating Equations (\ref{e:DJphi}) and (\ref{e:DJphiacf}) gives two values for $\theta_1$
\begin{equation}\label{e:thetanoexp}
\theta_1 = {1 \over 2} \left[- \left({1\over \phi_1} + \phi_1 \right) \pm \sqrt{{1 \over \phi_1^2} +  \phi_1^2 -2 + {4 (1 - \phi_1^2) \tau^2 P^2 \over \sigma_1^2}} \right]
\end{equation}
Taking the positive root we find an excellent match to the data. For values of $e > e_{\rm crit}$ the first term $ {1 \over 2} ({1\over \phi_1} + \phi_1) \approx 1$ and recognizing the Taylor expansion of $- \exp (- x) \approx x - 1$ we simplify this expression to
\begin{equation}\label{e:thetath}
\theta_1 = - \exp{ \left( - {f_\theta \over 2}\sqrt{{1 \over \phi_1^2} +  \phi_1^2 -2 + {4 (1 - \phi_1^2) \tau^2 P^2 \over \sigma_1^2}}\right) }
\end{equation}
where $f_\theta$ is a fourth free parameter. \\

We summarize this section by clarifying that we have four free parameters in our ARMA model ($f_\theta$, $f_\phi$, $f_\sigma$, $k$) fully determinable by our $N$-body simulations, with which we calibrate our model for use in MC simulations.

\section{Results: ARMA analysis of the $N$-body simulations}\label{s:N}

We now return to the $N$-body simulations, and use the time series of angular momenta that are generated to calibrate the ARMA model. These parameters fully define both the RR and NR processes. Once they are known, they can be used to generate new angular momentum series that have the correct statistical properties, much like what is done in regular MC simulations. We will exploit this method in Section \ref{s:MC}.

When a test star has mass, there will be a back-reaction on the field stars known as resonant friction, analogous to dynamical friction \citep{Rau96}. In our simulations, we consider both the case that the test star is massless and that it has the same mass as the field stars to calibrate the model parameters. Differences in the parameters then result from resonant friction.

We use the ``R'' language for statistical computing to calculate the ARMA model parameters from our simulations. We input the individual angular momentum time series of a test star, and use a maximum likelihood method to calculate $\phi_1$, $\theta_1$ and $\sigma_1$. We find that the drift term $\langle \Delta J \rangle$ for stars of all eccentricities is $\mathcal{O}(10^{-6})$. \\

\begin{figure}
\begin{center}
			\includegraphics[height=85 mm,angle=270 ]{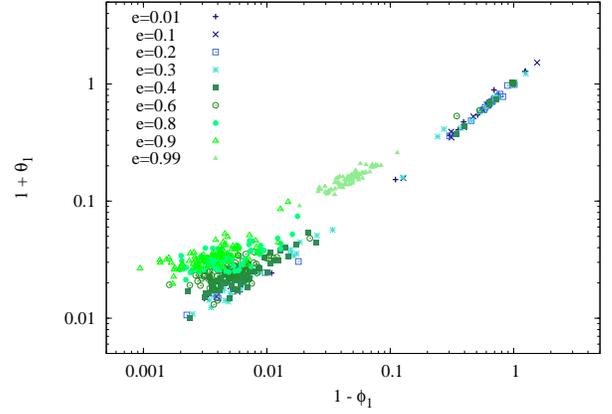}
		\caption{Scatter plot of ($1-\phi_1$) vs ($1+\theta_1$) as found from data-analysis on the $N$-body runs, for the case of massless test stars. For most cases, the parameters cluster, leading to a concentration of points around small values; reasons for exceptions are discussed in the text. From inspection of several individual cases, we see that the test stars of intermediate eccentricities which have values $\phi_1, |\theta_1| \lesssim 1$ (to the bottom left of plot) have non-zero autocorrelations for long times, similar to Figure \ref{f:acf}. This is as expected from Equation (\ref{e:acf}). Test stars with $\phi_1, \theta_1 \approx 0$ have ACFs that are close to zero everywhere, even for short times.
		\label{f:f_vs_t}}
\end{center}
\end{figure}

\begin{figure}
\begin{center}
			\includegraphics[height=85 mm,angle=270 ]{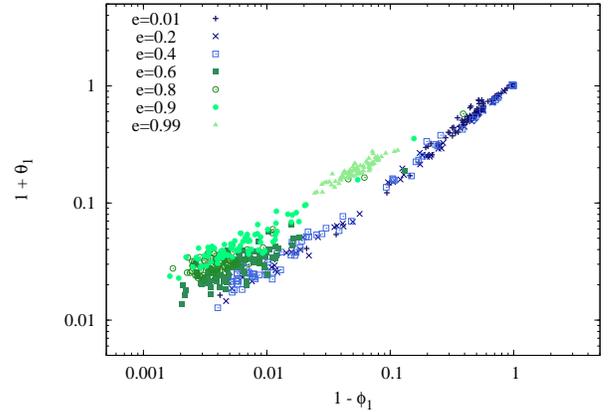}
		\caption{Scatter plot of ($1-\phi_1$) vs ($1+\theta_1$) as found from data analysis on the $N$-body runs, for the case of massive ($m = 10 \Mo$) test stars. We find that, for most eccentricities, the values for $1-\phi_1$ and $1+\theta_1$ cluster around those for the massless cases. 
		\label{f:M_f_vs_t}}
		\end{center}
\end{figure}

In Figure \ref{f:f_vs_t} and \ref{f:M_f_vs_t} we show scatter plots of the model parameters $1-\phi_1$ and $1+\theta_1$, for simulations in which the test stars are massless and massive respectively. We present these quantities rather than $\phi_1$ and $\theta_1$ themselves because they span several orders of magnitude and, physically speaking, the relevant question is by how much the parameters differ from (minus) unity. For intermediate eccentricities ($0.4 \lesssim e \lesssim 0.9$), $1-\phi_1$ is much smaller than unity, indicating that the angular momentum autocorrelations are indeed persistent, and there are significant torques between stellar orbits. However, the coherent torques are mixed with NR noise (the classical NR as treated in \citet{Bin08}), and as a result $1+\theta_1\ll1$ as well. These values for $1-\phi_1$ and $1+\theta_1$ confirm that although the ACF for RR is finite for long times, it is much smaller than unity --- which would be the case if there was no NR noise at all. 

For very small and very large eccentricities, we find $\phi_1\approx\theta_1\approx0$, so there are no persistent torques of significance. The reason for the absence of coherent torques on the high and low eccentricity limits are different: for very small eccentricities, the secular torque on a stellar orbit approaches zero as $\tau\propto e$ \citep{Gur07}. At the high eccentricity end, there are significant torques, but the coherence time is so short due to general relativity, that the persistence of these torques is negligible, and there is no coherent effect \citep[quenching of RR: see][]{Mer10b}. Test stars with eccentricities of $0.2 \le e \le 0.4$ have a large scatter in their $1 - \phi_1$ and $1+\theta_1$ values. This is due to their proximity to $e_{\rm crit}$, the eccentricity at which NR and RR compete for dominance at this radius in our model.

\begin{figure}
\begin{center}

			\includegraphics[height=85 mm,angle=270 ]{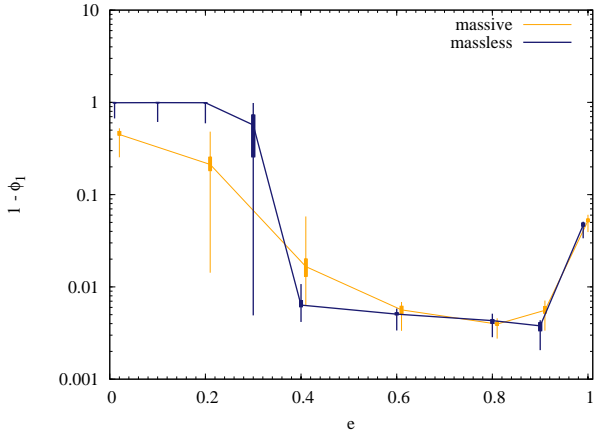}
		\caption{Median value of $1-\phi_1$, plotted on a log scale, for both massless and massive runs, computed for 80 test stars in each eccentricity bin. Closed boxes denote the 45th and 55th percentiles, while the lines indicate the one sigma values. Note that for very high eccentricities the values increase, which in the interpretation of Section \ref{ss:theory} means that the coherence time has decreased. This is consistent with the fact that for $e \ge 0.92$, GR precession becomes dominant over mass precession. The deviation at low eccentricities between values for massless and massive particles may due to resonant friction. \label{f:median_phi}}
		\end{center}
\end{figure}

\begin{figure}
\begin{center}

			\includegraphics[height=85 mm,angle=270 ]{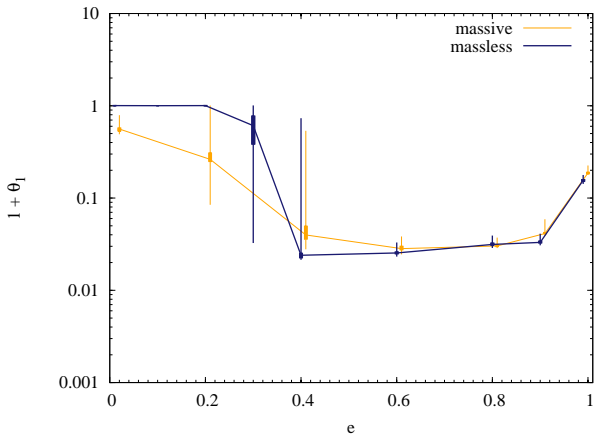}
		\caption{Median value of $1+\theta_1$, plotted on a log scale, for massless and massive runs, computed for 80 test stars in each eccentricity bin. Closed boxes denote the 45th and 55th percentiles, while the lines indicate the one sigma values. The results for massless and massive particles follow each other closely except at the low-eccentricity end where there is a large scatter in the data. \label{f:median_theta}}
		\end{center}
\end{figure}

\begin{figure} [ht]
\begin{center}
			\includegraphics[height=85 mm,angle=270 ]{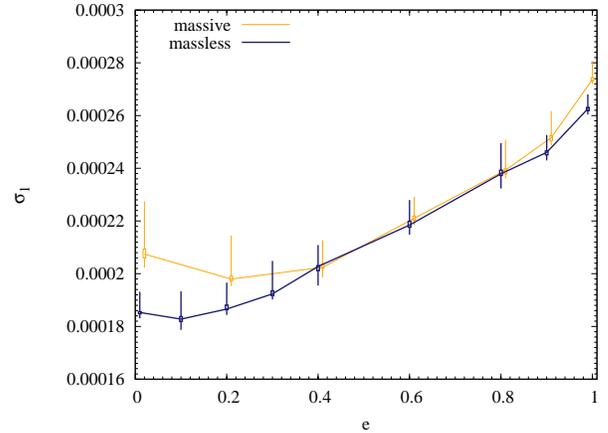}
		\caption{Data for the model parameter $\sigma_1$ as a function of eccentricity for massless and massive test stars. Plotted are the 45th to 55th percentiles (open boxes), one sigma values (lines), and 50th percentile for each eccentricity. The results for massive test stars are shifted along the $x$-axis for clarity. The values follow each other closely with the exception of low eccentricities. \label{f:sigma_2m}}
\end{center}
\end{figure}

In Figures \ref{f:median_phi} and \ref{f:median_theta} we compare the median values of $1 - \phi_1$ and $1 + \theta_1$ for both massive and massless test stars. There is significant scatter in the data near $0.2 \le e \le 0.4$, where NR relaxation competes with RR, as can be seen from the 45th, 55th and one sigma ($34.1 - 84.1$) percentiles. The differences between the median values for massless and massive test stars are very small, such that we did not see a strong feedback effect due to resonant friction, except for small stellar eccentricities. Resonant friction appears to drag stars away from circular angular momentum. The same deviation is seen for the $\sigma_1$ model parameter in Figure \ref{f:sigma_2m}.

\begin{table}
\caption{Median values of $1-\phi_1$, $1+\theta_1$ and $\sigma_1$ for massless test stars as a function of eccentricity.}
\begin{center}
\begin{tabular}{cccc}
\hline
\hline
$e$       & $1-\phi_1$	 & $1+\theta_1$ & $\sigma_1$ ($10^{-4}$) \tabularnewline
\hline
0.01 & 0.992 & 1.004 & 1.853 \tabularnewline
0.1   & 0.993 & 1.002 & 1.828 \tabularnewline
0.2   & 0.991 & 1.003 & 1.866 \tabularnewline
0.3   & 0.569 & 0.609 & 1.923 \tabularnewline
0.4	& 0.006 & 0.024 & 2.027 \tabularnewline
0.6	& 0.005 & 0.025 & 2.184 \tabularnewline
0.8	& 0.004 & 0.031 & 2.379 \tabularnewline
0.9	& 0.004 & 0.033 & 2.460 \tabularnewline
0.99	& 0.047 & 0.156  & 2.626 \tabularnewline
\hline
\end{tabular}
\end{center}
\label{t:epts}
\end{table}

From here on, we use only the results of the massless simulations as we have a larger range of test star eccentricities and have found the ARMA parameters ($\phi_1, \theta_1, \sigma_1$) to be similar in value. We refer the reader to Table \ref{t:epts} for exact quantities.

The ARMA model parameters for both very low eccentricities ($e \sim 0.01$) and high eccentricities ($e \gtrsim 0.4$) tend to cluster together. For these cases one form of relaxation, either NR or RR, dominates the form of the ACFs, and taking mean values of $(\phi_1, \theta_1)$ gives an accurate representation of the population. However, near the $e_{\rm crit}$ boundary where the two relaxation effects compete for dominance, $(\phi_1, \theta_1)$ do not consistently cluster but rather choose NR or RR values which can vary significantly. Hence we choose the median value in each case to compare with our theory.

\begin{figure}
\begin{center}
			\includegraphics[height=85 mm,angle=270 ]{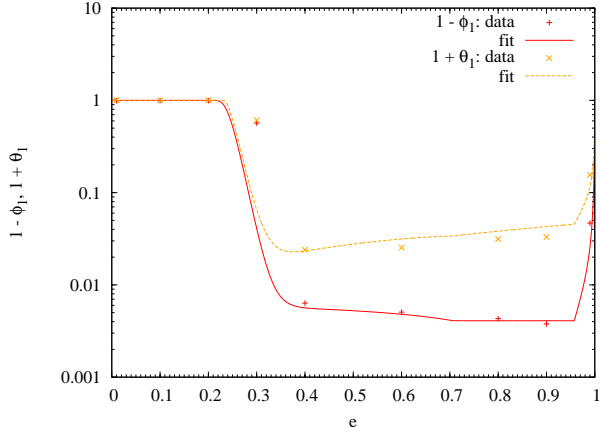}
		\caption{Median value of $(1-\phi_1)$ and $(1+\theta_1)$, compared with the theoretical values from (\ref{e:phith}, \ref{e:thetath}). There is a sharp transition between eccentricities $0.3 \lesssim e \lesssim 0.4$ where the two effects of NR and RR compete. A second transition occurs at high eccentricities ($e > 0.92$) as the coherence time, and hence $\phi_1$, decreases due to general relativistic effects. The theory does not exactly match the median value of the data at $0.2 < e < 0.3$. However we can vary the value of $e_{\rm crit}$ to reconcile this difference and we find that this is not important for our results in the next section.
	\label{f:th}}
	\end{center}
\end{figure}

\begin{figure} [ht]
\begin{center}
			\includegraphics[height=85 mm,angle=270 ]{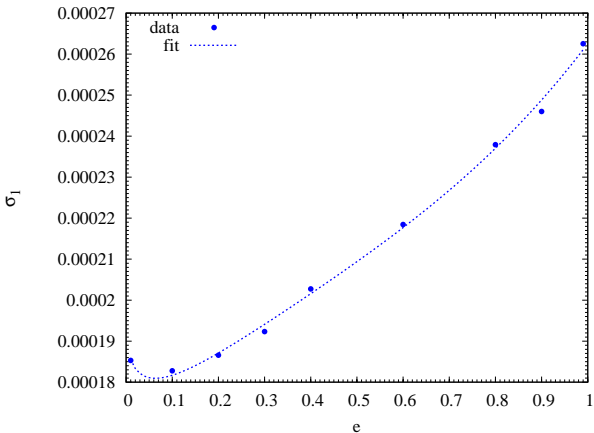}
		\caption{Median values of the model parameter $\sigma_1$ (related to non-resonant relaxation) and theoretical fit, as a function of eccentricity for massless test stars. \label{f:sigma}}
		\end{center}
\end{figure}

In Figures \ref{f:th} and \ref{f:sigma} we compare the experimental median values of $(1-\phi_1)$, $(1+\theta_1)$ and $\sigma_1$ to our theoretical model. We find that good agreement between the $N$-body simulations and the model is obtained by assuming values for the free parameters $(f_\phi, f_\theta, k) = (0.105, 1.2, 30)$. The value of $f_\phi$ shows that the coherence time $t_{\phi}$ is much less than the precession time $t_{\rm prec}$; see Equation (\ref{e:tphi}). Fitting $f_{\sigma}$ as a function of $e$ we find
\begin{equation}\label{e:fsigma}
f_{\sigma} = 0.52 + 0.62 e - 0.36 e^2 + 0.21 e^3 -0.29  \sqrt{e}. 
\end{equation}

\section{Monte Carlo simulations and applications}\label{s:MC}

Our aims in this section are to (1) explore the long-term evolution of stars around an MBH, (2) investigate the depletion of stars around an MBH due to RR in context of the ``hole'' in late-type stars in the GC, and (3) study the evolution of the orbits of possible S-star progenitors from different initial setups. Direct $N$-body simulations of secular dynamics are challenging due to the inherent large range in timescales. The precession time is orders of magnitude larger than the orbital time, and in order to study relaxation to a steady-state, the system needs to evolve for a large number of precession times. Currently it is not feasible to do this using $N$-body techniques.
Instead we draw on MC simulations to study long term secular evolution. For this purpose, we have developed a two-dimensional (energy and angular momentum) MC code.  The main new aspect compared to earlier work is that it implicitly includes evolution due to RR. The angular momentum diffusion is based on the ARMA(1, 1) model presented in Section \ref{s:stat}. For the energy evolution we use a similar method as in \citet{Hop09b}, which was in turn based on \citet{Sha78}. A key feature of this method is a cloning scheme, which allows one to resolve the distribution function over many orders of magnitude in energy, even though the number of particles per unit energy is typically a steep power-law, $dN/dE\propto E^{-9/4}$. We will not describe the cloning scheme here; for details we refer the reader to \citet{Sha78} and \citet{Hop09b}. 

We base our model of a galactic nucleus on parameters relevant for the GC; see Section \ref{s:nbody_sims}. We set the MBH mass as $\Mbh=4\times10^6\Mo$, which affects the physics through the GR precession rate and the loss cone. We linearize the system by considering the evolution of test stars on a fixed background stellar potential. We assume a single-mass distribution of stars of either $m=1 \Mo$ or $m=10\Mo$, with the number of stars within a radius $r$, $N(< r)$, following a power-law profile as in Equation (\ref{e:N_h}). 

\begin{figure}
		\includegraphics[height=85 mm,angle=270 ]{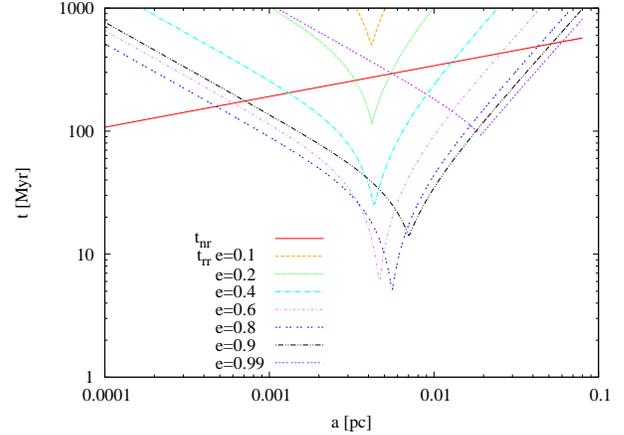}
		\caption{RR and NR timescales in units of Myr as a function of semi-major axis $a$ in our Galactic nucleus model for different eccentricities. 
		\label{f:trr_tnr_e_a}}
\end{figure}

\begin{figure}
\begin{center}
			\includegraphics[height=85 mm,angle=270 ]{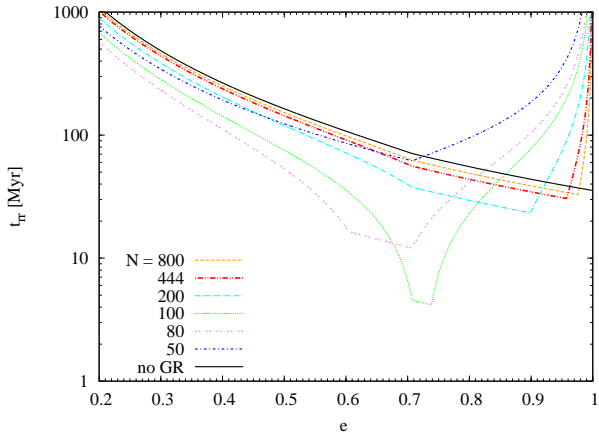}
		\caption{Resonant relaxation timescale $t_{\rm rr}$ for varying stellar mass ($N m$ where $m = 10 \Mo$) within $0.01 \pc$ as a function of $e$. In this plot the density power-law index $\alpha = 1.75$. The $e$ at which RR is most efficient decreases with decreasing mass. The solid black line indicates the value of $t_{\rm rr}$ if the effects of GR precession are not taken into account (and is independent of number $N$). In our fiducial model the optimal efficiency (shortest $t_{\rm rr}$) is for $N m = 1000 \Mo$. At this radius, GR effects are important for RR and $t_{\rm rr}$ is not independent of $N$.
		\label{f:trr_e_fn_M}}
		\end{center}
\end{figure}

In Figure \ref{f:trr_tnr_e_a} we plot the NR and RR timescales for stars with different orbital eccentricities as a function of semi-major axis for this model with $m = 10 \Mo$. A cuspy minimum arises in the RR timescales where GR precession cancels with Newtonian mass precession. In their GC model \citet{Koc10} find, using \citet{Eil09} parameters, the cuspy minimum of RR near $r = 0.007 \pc$ (see their Figure 1), which falls precisely within the range of values in our model ($0.004 - 0.02 \pc$). Figure \ref{f:trr_e_fn_M} shows the RR time as a function of eccentricity $e$ for different stellar masses within $0.01 \pc$. Models with different enclosed masses will be used in Section \ref{ss:sstars}.

In reality, several stellar populations are present in the GC, and our choice for a single mass model is therefore a simplification. The slope $\alpha=1.75$ is smaller than would be expected for strong mass-segregation \citep{Ale09, Kes09}, but was chosen to make the model self-consistent as a collection of interacting single mass particles will relax to have this distribution \citep{Bah76}. 

\subsection{Angular momentum} 
For the angular momentum evolution of a test star, we go through the following steps: we first calculate the ARMA model parameters for one orbital period $(\phi_1, \theta_1, \sigma_1)$ through Equations ((\ref{e:phith}), (\ref{e:thetath}), (\ref{e:sigmath})). We then find the model parameters $(\phi_N, \theta_N, \sigma_N)$ for the time step $\delta t$, where $N (=  \delta t / P) \in \mathbb{R} > 0 $, using Equations (\ref{e:pt}) - (\ref{e:sigmaN}). Finally, the step in angular momentum is given by Equation (\ref{e:DJN})
\begin{equation}\label{e:DJN_repeat}
\begin{split}
\Delta_N J_t &= \phi_N\Delta_N J_{t-N} + \theta_N\epsilon_{t-N}^{(N)} + \epsilon^{(N)}_t ;\\
\langle [\epsilon^{(N)}]^2\rangle &= \sigma_N^2.
\end{split}
\end{equation}

\subsection{Energy} 
The energy of a star is given in units of the square of the velocity dispersion at the radius of influence, $v_h^2$, such that $E = r_h / 2 a$.
The step in energy during a time $\delta t$ is
\begin{equation}\label{e:DyyBW}
\Delta E = \xi E \left({\delta t\over t_{\rm NR}}\right)^{1/2},
\end{equation}
where $\xi$ is an independent normal random variable with zero mean and unit variance, and $t_{\rm NR}$ is given by Equation (\ref{e:tNR}) with $A_{\rm NR} = 0.26$.

\subsection{Initial conditions and boundary conditions} 
For our steady-state simulations, we initialize stars with energies $E=1$ and angular momenta drawn from a thermal distribution, $dN(J)/dJ\propto J$. We follow the dynamics of stars between energy boundaries $0.5=E_{\rm min} < E < E_{\rm max} = 10^4$ (for our model this corresponds to $2.31 \pc > a > 1.155 \times 10^{-4} \pc$). Stars that diffuse to $E<E_{\rm min}$ are reinitialized, but their time is not set to zero \citep[zero-flow solution;][]{Bah76, Bah77}. Stars with $E>E_{\rm max}$ are disrupted by the MBH and also reinitialized while keeping their time. Stars have angular momenta in the range $J_{\rm lc}<J<1$, where $J_{\rm lc}$ is the angular momentum of the last stable orbit. Stars with $J<J_{\rm lc}$ are disrupted and reinitialized, while stars with $J>1$ are reflected such that $J\to2-J$.

\subsection{Time steps} 

In the coherent regime, the change in angular momentum during a time $\delta t$ is of order $\delta J \sim \tau\delta t \sim A_{\tau} e (m/\Mbh) \sqrt{N_<} (\delta t/P)$. The changes in angular momentum should be small compared to either the size of the loss cone $J_{\rm lc}$, or the separation between $J$ and the loss cone, $J - J_{\rm lc}$; and compared to the distance between the circular angular momentum  $J_c$ and $J$, i.e., compared to $1-J$. We therefore define the time step to be
\begin{equation}\label{e:dt}
\delta t  = f_t ~ \tau(a,e)^{-1} \min\left[J_c - J, \max\left(|J-J_{\rm lc}|, J_{\rm lc}\right)\right],
\end{equation}
where $f_t \ll1$. After some experimenting we found that the simulations converge for $f_t\leq 10^{-3}$, which is the value we then used in our simulations.

\subsection{Treatment of the loss cone} 

Stars with mass $m$ and radius $\rs$ on orbits which pass through the tidal radius
\begin{equation}\label{e:rt}
r_t = \left( {2 \Mbh \over m }\right)^{1/3} \rs
\end{equation}
are disrupted by the MBH. The loss cone is the region in angular momentum space which delineates these orbits and is given by
\begin{equation}\label{e:Jlc}
J_{\rm lc} = \sqrt{2 G \Mbh r_t}.
\end{equation}
Our prescription for the time step, which is similar to that used in \citet{Sha78}, ensures that stars always diffuse into the loss cone and cannot jump over it. Stars are disrupted by the MBH when their orbital angular momentum is smaller than the loss cone, {\it and} they pass through periapsis. Only if ${\rm floor}(t/P)-{\rm floor}\left[\left(t-\delta t\right)/P\right]>0$, do we consider the star to have crossed periapsis during the last time step and hence to be tidally disrupted (or directly consumed by the MBH in case of a compact remnant). In that case we record the energy at which the star was disrupted and initialize a new star, which starts at the time $t$ at which the star was disrupted. 

\section{Results}\label{s:Results}

\subsection{Evolution to steady-state}\label{ss:ss}

We first consider a theoretical benchmark problem, that of the steady-state distribution function of a single-mass population of stars around an MBH. With $m=1 \Mo$ this solution appears after $\sim 100 \Gyr$ ($10$ energy diffusion timescales; see Appendix \ref{s:app:energy}). In order to highlight the differences caused by RR relative to NR, we define the function
\begin{equation}\label{e:gBW}
g(E, J^2) \equiv {E^{5/4}\over J^2}{d^2 N(E, J^2)\over d\ln E d\ln J^2}.
\end{equation}
For a \citet{Bah76} distribution without RR, this function is constant. In \citet{Too09}, simulations similar to those presented here were used, except that angular momentum relaxation was assumed to be NR. It was shown that indeed $g(E, J^2)$ is approximately constant for all $(E, J)$ for the NR case; see Figure A1 in that paper. 

\begin{figure}
\begin{center}
			\includegraphics[height=90 mm,angle=270 ,trim=2.5cm 0cm 0cm 0cm,clip]{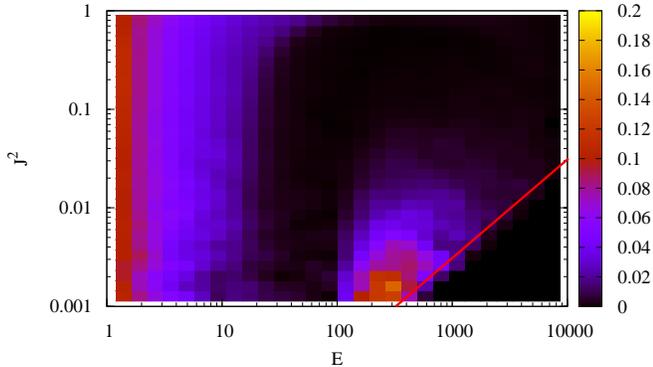}
		\caption{Two-dimensional distribution of stars, normalized such that for an isothermal \citet{Bah76} profile the distribution should be constant (see Equation (\ref{e:gBW})). Note that angular momentum is in units of the circular angular momentum, so that the loss cone (indicated by the red line) is not constant. At high energies it is clear that there is a depletion of stars, and that the distribution of angular momenta is far from isotropic.\label{f:fEJ}}
		\end{center}
\end{figure}

In Figure \ref{f:fEJ} we show a density plot of the function $g(E, J^2)$ from our steady-state simulations, which illustrates several interesting effects of RR. At low energies, $g(E, J^2)$ is constant which is to be expected since the precession time is of similar magnitude as the orbital period, such that there are no coherent torques. At higher energies ($E \gtrsim 10$), there are far fewer stars than for a classical \citet{Bah76} cusp. This is a consequence of enhanced angular momentum relaxation, where stars are driven to the loss cone at a higher rate than can be replenished by energy diffusion, and was anticipated by one-dimensional Fokker-Planck calculations \citep[][see their Figure 2]{Hop06a}. At the highest energies considered ($E > 100$) most orbits accumulate close to the loss cone, and there are a few orbits populated at larger angular momenta, with a ``desert'' in between.

\begin{figure}
\begin{center}
			\includegraphics[height=85 mm,angle=270 ]{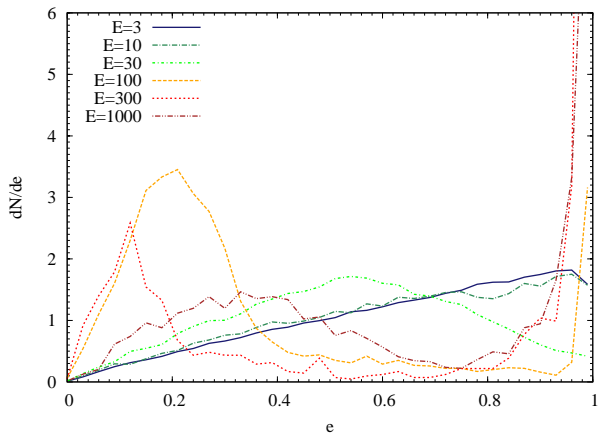}
		\caption{Eccentricity distribution at fixed energies. For high energies the distribution of eccentricities is bimodal due to RR, whereas for low energies the distribution is isotropic with a cutoff at the loss cone.\label{f:ecc}}
\end{center}
\end{figure}

To further illustrate the reaction of the stellar orbits to RR, in Figure \ref{f:ecc} we show the distribution of eccentricities of stars in slices of energy space. As in Figure \ref{f:fEJ} we see that for higher energies the distribution is double peaked, with most stars accumulating at the highest eccentricities, and another peak at eccentricities around $e = 0.2$. This distribution can be understood as follows: RR is very effective at intermediate eccentricities, where torques are strong and the coherence time is very long. As a result, the evolution of such eccentricities occurs on a short timescale. At high eccentricities, the coherence time is short due to GR precession, while at low eccentricities the torques are very weak. At such eccentricities, evolution is very slow. The eccentricity therefore tends to ``stick'' at those values, leading to the two peaks in the distribution. It is interesting to note that $N$-body simulations by \citet{Rau98} also show a rise of the angular momentum distribution near the loss cone. This result however may be affected by the new dynamical mechanism recently described by \citet{Mer11}.

\begin{figure}
\begin{center}
			\includegraphics[height=85 mm,angle=270 ]{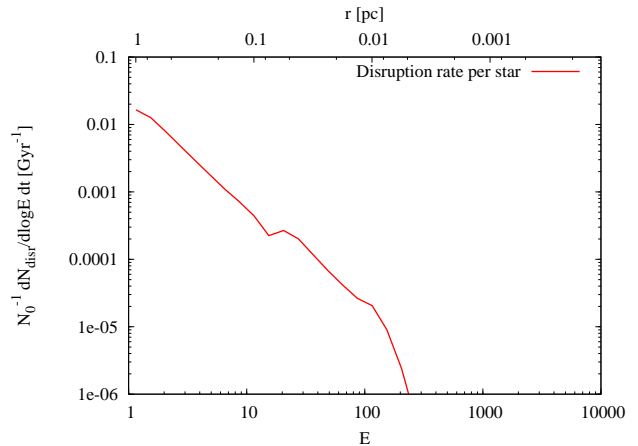}
		\caption{Disruption rate per star per $\log E$, in units of ${\rm Gyr}^{-1}$, as a function of energy. The rate drops quickly as the stellar distribution is depleted due to RR. \label{f:disr}}
		\end{center}
\end{figure}

In Figure \ref{f:disr} we show the rate of direct captures of stars by the MBH per log energy. Loss cone theory \citep{Fra76, Lig77, Coh78, Sye99, Mag99} yields that the disruption rate continues to rise with decreasing energy for energies larger than the ``critical energy\footnote{The critical energy is approximately the energy where the change in angular momentum per orbit equals the size of the loss cone.}'', which in our case is at $E<1$. This is in accordance with Figure \ref{f:disr}. As the figure shows, the rate drops quickly; this is due to the depletion of stars, and was also found (though to a lesser extent, compare their Figure 4) in \citet{Hop06a}.

A limiting factor of our approach in this section is that the potential of our Galactic nucleus does not evolve; hence we cannot look at collective effects (e.g., instabilities). Ideally an iterative process should be used to self-consistently find the steady-state solution; we have shown that RR significantly depletes stars at inner radii. To what extent this will affect our steady-state solution is beyond the scope of the paper. We note the RR time does not formally depend on the stellar number density until such energies and angular momenta that GR precession becomes important. 

\subsection{A Depression in the Galactic Center}\label{ss:GC_hole}

Early studies of integrated starlight at the GC detected a dip in the CO absorption strength within $15 \arcsec$ of the MBH, Sgr A* \citep{Sel90, Hal96}, which indicated a decrease in the density of old stars \citep[see also][]{Gen96, Gen03a, Fig00, Fig03b, Sch07, Zhu08}. Recent observations by \citet{Do09}, \citet{Buc09}, and \citet{Bar10} have confirmed this suggestion and, in particular, revealed that the distribution of the late-type stars in the central region of the Galaxy is very different than expected for a relaxed density cusp around an MBH.

\citet{Do09} use the number density profile of late-type giants to examine the structure of the nuclear star cluster in the innermost $0.16 \pc$ of the GC. They find the surface stellar number density profile, $\Sigma (R) \propto R^{-\Gamma}$, is flat with $\Gamma = 0.26 \pm 0.24$ and rule out all values of $\alpha > 1.0$ at a confidence level of $99.7\%$. The slope measurement cannot constrain whether there is a hole in the stellar distribution or simply a very shallow power law volume density profile. \citet{Buc09} show that the late-type density function is flat out to $10 \arcsec$ with $\Gamma = -0.7 \pm 0.09$ and inverts in the inner $6 \arcsec$, with power-law slope $\Gamma = 0.17 \pm 0.09$. They conclude that the stellar population in the innermost $\sim 0.2 \pc$ is depleted of both bright and faint giants. These results are confirmed by \citet{Bar10} who find that late-type stars with $m_k \le 15.5$ have a flat distribution inside of $10 \arcsec$. 

This result has changed our perception of the nuclear star cluster at the center of our Galaxy. Stellar cusp formation theory predicts a volume density profile $n(r) \propto r^{-\alpha}$ of relaxed stars with $\alpha = 7/4$ for a single mass population \citep{Bah76} and has been confirmed in many theoretical papers with different methods, including Fokker-Planck \citep[e.g.,][]{Coh78,Mur91}, MC \citep[e.g.,][]{Sha78, Fre02}, and $N$-body methods \citep{Pre04, Bau04a, Pre10}. Theory also predicts that a multi-mass stellar cluster will mass segregate to a differential distribution with the more massive populations being more centrally concentrated. The indices of the power-law density profiles for the different mass populations can vary between $3/2 < \alpha < 11/4$ \citep{Bah77,Fre06,Ale09}. 

There have been two types of solutions put forward in the literature to explain the discrepancy between cusp formation theory and observations of the old stellar population in the GC. The first maintains that the stellar cusp in the GC is relaxed as predicted, but additional physical mechanisms have depleted the old stars in the central region. Such mechanisms include strong mass segregation \citep{Ale09} and envelope destruction of giants by stellar collisions \citep[see][and references within]{Dal09}. The second solution proposes an unrelaxed cusp, either by the depletion of the stars due to the infall of an intermediate-mass black hole \citep[IMBH;][]{Lev06, Bau06},  or from a binary black hole merger \citep{Mer06,Mer10}. Kinematical information for the old stellar population in the GC however shows a lack of evidence for any large-scale disturbance of the old stellar cluster \citep{Tri08}, and \citet{she10} find that the Galaxy shows no observational sign that it suffered a major merger in the past $9-10 \Gyr$. In addition to this, the dynamics of the S-stars and of the MBH greatly restricts the parameter space available for the presence of an IMBH in the GC \citep{Gua09}. For a summary see the latest review by \citet{Gen10}.

Here we propose that these observations can be explained in part by a depression of stars carved out by RR acting together with tidal disruption. The underlying reason is that stars move rapidly in angular momentum space and enter the loss cone at a higher rate than are replaced by energy diffusion. We show that when we initialize the system as a cusp that, even though steady-state is reached in longer than a Hubble time, the depression forms on a shorter timescale. 

\begin{figure}
\begin{center}
			\includegraphics[height=85 mm,angle=270 ]{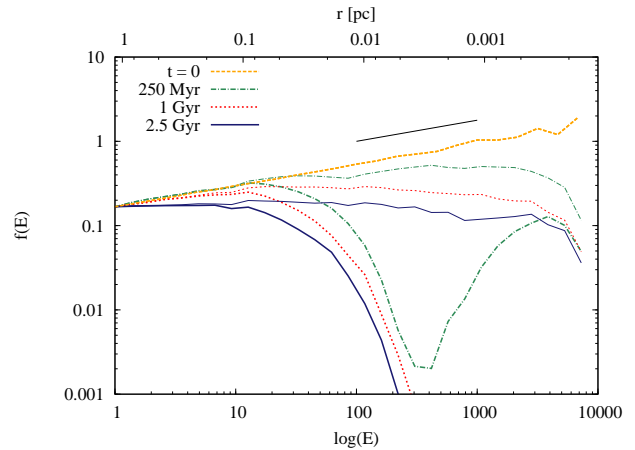}
\caption{Carving out of a depression in the distribution of stars around an MBH for simulations in which $m=10\Mo$. The broken orange line shows the stellar distribution at the start of the simulation (the solid black line above indicates a slope of $1/4$). The thick colored lines show the results for RR while the corresponding (in color and style) thin lines show results for simulation in which RR is switched off. The energy, or equivalently radius, at which RR becomes efficient at carving out the depression is clearly seen on the scale of $\sim 0.1 \pc$. 
		\label{f:f_E}}
		\end{center}
\end{figure}

We perform simulations of stars, both for $m = 1 \Mo$ and $m = 10 \Mo$, initially distributed in a Bahcall-Wolf (BW) cusp around the MBH, such that $dN/dE \propto E^{-9/4}$. We present first the results for  $m = 10 \Mo$, chosen to reflect the dominant stellar population at small radii. We plot the stellar distribution function $f(E)$, which scales as $E^{1/4}$ for BW cusp, at various times during the simulation as the stars move in angular momentum space due to RR; see Figure \ref{f:f_E}. In $\sim 1 \Gyr$ a depression is carved out of the population of stars, significantly depleting the stellar population around the MBH out to $\sim 0.2 \pc$. To conclusively demonstrate that RR is the reason for this depression, we also show the distribution of stars for simulations in which we have switched RR off (by setting $\phi_1 = \theta_1 = 0$, $\sigma_1 \ne 0$) such that angular momentum relaxation follows a random walk. In this case no depression forms. These simulations show that, as the cusp is destroyed on timescales of a few $\Gyr$ due to RR, a BW cusp is not a solution to the distribution of stars around an MBH. We see the same appearance of a depression when we substitute our non-resonant parameters (both energy and angular momentum) for those of \citet{Eil09}.

\begin{figure}
\begin{center}
			\includegraphics[height=85 mm,angle=270 ]{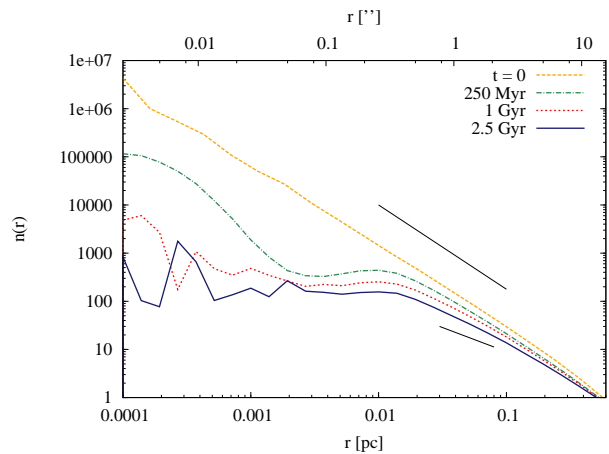}
\caption{Number density of stars as a function of radius (in parsec and arcsecond) from the MBH for $m=10\Mo$ simulations. The stellar distribution begins to deviate from the BW solution at $\sim 0.5 \pc$. The black lines indicate a slope of $-1.75$ (top) and $-1.0$ (bottom right).
		\label{f:n_r}}
		\end{center}
\end{figure}

\begin{figure}
\begin{center}
			\includegraphics[height=85 mm,angle=270 ]{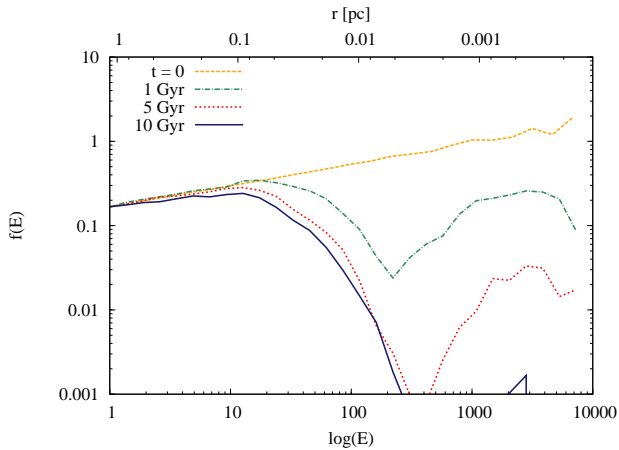}
\caption{Carving out of a depression in the distribution of stars around an MBH for simulations in which $m=1\Mo$. The timescale on which the depression develops is longer than for simulations in which the potential is built up of $10 \Mo$ stars. Stars which have an energy $E \gtrsim 1000$ survive longer than those with $100 \gtrsim E \gtrsim 1000$ as RR is ineffective so close to the MBH due to GR.
		\label{f:f_Em1}}
		\end{center}
		
\end{figure}

\begin{figure}
\begin{center}
			\includegraphics[height=85 mm,angle=270 ]{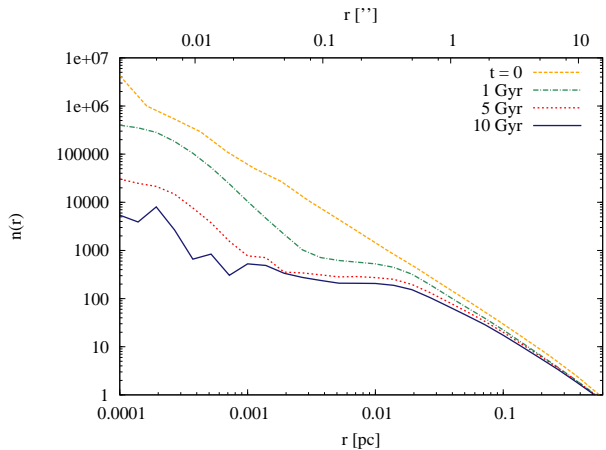}
\caption{Number density of stars as a function of radius (in parsec and arcsecond) from the MBH for $m=1\Mo$ simulations. The depression in stars takes an order of magnitude longer in time to develop than the $m= 10 \Mo$ simulations.
		\label{f:n_rm1}}
		\end{center}
\end{figure}

In Figure \ref{f:n_r}, we plot the number density of stars as a function of radius from the MBH. We do this by sampling each stellar orbit randomly in mean anomaly. Black lines indicate the initial BW number density slope of $-1.75$ and a slope of $-1$. The slope begins to deviate, albeit gently, from the BW solution at $\sim 0.5 \pc$ and decreases to $-1$ at $\sim 0.1 \pc$. We present results from $m=1 \Mo$ simulations in Figures \ref{f:f_Em1} and \ref{f:n_rm1} even though, due to mass segregation, we do not expect low-mass stars to dominate the potential so close to an MBH. The major difference between the two simulations is the timescale on which the depression forms, $1 \Gyr$ and $10 \Gyr$, respectively.


To reproduce the radial extent of the depletion of old stars in the observations of \citet{Do09}, \citet{Buc09}, and \citet{Bar10}, the size of the hole in energy space must be significantly larger, by a factor of $5-10$, than what we find in our simulations. In addition, a hole in energy space alone cannot give rise to an inverted profile in surface number density. We stress however that the potential in our model does not evolve in response to RR. As the depression develops and the number of stars at small radii decreases, the energy relaxation timescale, $t_{\rm NR} \propto N^{-1}$, will increase. As $N$ is not important for the RR timescale for all energies but the largest (GR precession at small radii affects the precession rate), this could further enlarge the depression as the rate of stars flowing inward toward the MBH decreases. \\

\begin{figure}
\begin{center}
			\includegraphics[height=85 mm,angle=270 ]{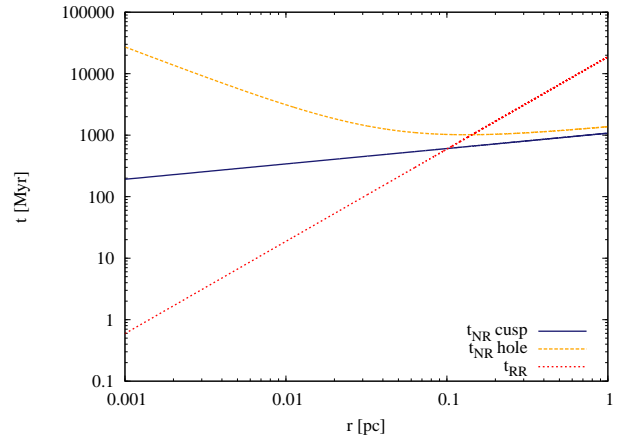}
\caption{NR (energy) timescale changing in response to the stellar potential ($m = 10 \Mo$) as it develops from a BW cusp to one with a hole in energy space (from $t_{\rm NR}^{\rm cusp}$ to $t_{\rm NR}^{\rm hole})$. The RR timescale is plotted here without GR precession as $\Mbh/m P$ and, at large radii, does not depend on the number of stars. Hence this timescale should remain constant above $\sim 10^{-2} \pc$ as the NR timescale increases. The radius at which $t_{\rm RR}$ is equal to $t_{\rm NR}$ moves outwards in response to the changes and hence the outer radius of the hole in stars will increase.
		\label{f:tnr_eq_trr}}
		\end{center}
		
\end{figure}

In Figure \ref{f:tnr_eq_trr}, we demonstrate the increase in the NR timescale in response to the hole in stars. We use a fit to the stellar density profile at $1 \Gyr$ from Figure \ref{f:n_r} to recalculate $t_{\rm NR}$, normalizing the profile such that the mass in stars at the radius of influence is equal to that of the MBH. We plot the RR timescale without GR precession as relativistic effects are unimportant at radii greater than $\sim 10^{-2}$; see Figure \ref{f:trr_tnr_e_a}. As the outer extent of the hole at $\sim 10^{-1} \pc$ is at a much larger distance from the MBH, the RR timescale at this scale will not be affected by changes in $N$. The increase in the radius at which $t_{\rm NR} = t_{\rm RR}$ strongly suggests that the outer radius of the hole should increase in response to the change in stellar potential. \\

One implication of this depression is that we expect no gravitational wave bursts to be observed from the GC \citep{Hop07}. For the implications of the existence of a depletion of stars in the vicinity of an MBH on predicted event rates of gravitational wave inspirals of compact objects (EMRIs) and bursts, see \citet{Mer10}.

\subsection{Dynamical evolution of the S-stars}\label{ss:sstars}

At a distance of $\sim0.01\pc$ from the MBH in the GC, there is a cluster of young stars known as the S-stars. This cluster has been the subject of extensive observational and theoretical work, for which we refer the reader to \citet{Ale05} and \citet{Gen10}.  The stars are spectroscopically consistent with being late O- to early B-type dwarfs, implying that they have lifetimes that are limited to between $\lesssim6$ and $400\Myr$ \citep{Ghe03a, Eis05, Mar08}. The upper constraint comes from the limit on the age of a B-type main-sequence star, while the lower constraint is a maximum age limit on a specific star (S2/S0-2). One particularly interesting question concerns the origin of these stars. There is a consensus in the literature in that they cannot have formed at their current location due to the tidal field of the MBH. Several scenarios incorporating the formation of these stars further out from the MBH, and transport inward through dynamical processes, have been proposed \citep[e.g.][]{Gou03, Ale04, Lev07, Per07, Loc08, Loc09, Mad09, Mer09a,  Gri10}.

Many leading hypotheses involve binary disruptions \citep{Hil88, Hil91}. In this scenario a close encounter between a binary system and an MBH leads to an exchange interaction in which one of the stars is captured by the MBH, and the other becomes unbound and leaves the system with a high velocity. For capture, the periapsis $r_p$ of the binary's orbit (with  combined mass $m_{\rm bin}$ and semi-major axis $a_{\rm bin}$) must come within the tidal radius 
\begin{equation}\label{e:rt_bin}
r_t =  \left( { 2 \Mbh \over m_{\rm bin} }\right)^{1/3} a_{\rm bin}.
\end{equation}
The captured star will remain with a semi-major axis $a_{\rm cap}$ that scales as $a_{\rm cap}\sim(\Mbh/ m_{\rm bin} )^{2/3}a_{\rm bin}$, and its eccentricity can be approximated as $1-e \sim(m_{\rm bin}/\Mbh)^{1/3}$.

A unifying aspect of all proposed formation mechanisms is that none of them lead directly to orbital characteristics that are in agreement with those of the S-stars; in particular, the binary disruption mechanism leads to highly eccentric orbits. Since the local RR time is shorter than the maximum age of the S-stars, it has been suggested that following the arrival of the S-stars on tightly bound orbits to the MBH, a subsequent phase of RR has driven them to their current distribution \citep{Hop06a, Lev07}. RR evolution of S-stars has been studied by \citet{Per09b} by means of direct $N$-body simulations. They find that high-eccentricity orbits can evolve within a short time (20 Myr), to an eccentricity distribution that is statistically indistinguishable to that of the S-stars. They conclude that the S-star orbital parameters are consistent with a binary disruption formation scenario. However, these simulations do not include GR precession, which we have found to be important for the angular momentum evolution of stars at these radii.

\subsubsection{Simulations}

We use our ARMA code, which includes GR precession, to study the evolution of the S-star orbital parameters within the binary disruption context. We initialize our test stars on high eccentricity orbits, $e = 0.97$ ($m_{\rm bin} \sim 20 \Mo$), with initial semi-major axes corresponding to those reported by \citet{Gil09} for the S-stars. We assume a distance to SgrA* of $8.0 \kpc$ \citep{Eis03,Ghe08a,Gil09}. For this distance, one arcsecond corresponds to a distance of $0.0388 \pc$. We do not include S-stars that are associated with the disk of O/WR stars at larger radii for two reasons. The first is that they most likely originated from the disk with lower initial orbital eccentricities than we simulate here. Second, RR is not the most effective mechanism for angular momentum evolution at their radii; torques from stars at the inner edge of the disk will dominate. We run the simulation one hundred times for each test star to look statistically at the resulting distribution.

We simulate two different initial setups. The first, which we call the {\it continuous} scenario, assumes that the tidal disruption of binaries is an ongoing process, as is most likely in the case considered by \citet{Per07}, where individual binaries are scattered into the loss cone by massive perturbers. The same initial conditions also hold if low angular momentum binaries are generated as a result of triaxiality of the surrounding potential. For this study we start evolving the stellar orbits at times that are randomly distributed between $[0,t_{\rm max}]$, and continue the simulation until the time $t=t_{\rm max}$. In our second setup,  which we call the {\it burst} scenario, we consider a situation in which S-stars are formed at once on eccentric orbits at $t=0$, and follow their evolution until $t = t_{\rm max}$; this enables us to compare directly to the results of \citet{Per09b} in whose simulations all stars commence their evolution simultaneously. A burst of S-star formation, over a short period of a few$\Myr$, could for example result from an instability in an eccentric stellar disk \citep{Mad09}; we note that the instability is even more effective for a depressed cusp \citep{Mad10}.

We use two different models for the galactic nucleus. For each we re-compute the ARMA parameters $(\phi, \theta, \sigma)$ and $t_{\rm NR}$ for the new profile and re-run the MC simulations using Equations (\ref{e:DJN_repeat}) and (\ref{e:DyyBW}) with the updated values. {\it Model A} is similar to the galactic nucleus model discussed in Section \ref{s:nbody_sims}. The stellar cusp has a density profile $n(r) \propto r^{-\alpha}$ with $\alpha = 1.75$, and the number of stars is normalized with $N (< 0.01 \pc) = 222$ and $m=10$. We chose this model such that it faithfully reflects the dominant stellar content at a distance of $0.01\pc$, where RR is most important and where the S-stars are located. We use {\it Model B} to simulate the response of the S-stars to a depression in the GC, as found in the previous section. This model has a smaller density power-law index of $\alpha = 0.5$ (which is within the $90\%$ boundaries of the bootstrapped values for $\alpha$ in the GC as found by \citet{Mer10}) and $N (< 0.01 \pc) =148$. Less mass at small radii leads to slower mass precession, pushing to a lower eccentricity the value at which mass precession cancels with GR precession; see Figure \ref{f:trr_e_fn_M} for an illustration of this effect where $\alpha = 1.75$. Interestingly for all values of $N (< 0.01 \pc) \lesssim 120$, this eccentricity lies at $\sim 0.7$ a location in eccentricity space where we do not observe S-stars, which causes a flattening of the observed cumulative distribution. \\

\begin{table} [th]
\caption{S-star Parameters}
\begin{center}
\begin{tabular}{cccccc}
\hline
\hline
Model A & $t_{\rm max}$ (Myr)       &	 6 & 10 & 20 & 100 \tabularnewline
\hline
$p$$^a$& Burst$^d$ & 0.012 & 0.147 & 0.007&  5e-4\tabularnewline
& Continuous$^e$ & 2e-4 & 0.002 & 0.05 & 0.02 \tabularnewline
$f > 0.97$$^b$ & Burst     & .40 & .29 & .24 & .07 \tabularnewline
& Continuous             &  .47 & .40 &  .33& .16 \tabularnewline
DR$^c$ & Burst        & .23 & .32 & .46 & .72 \tabularnewline
& Continuous             & .09  & .17 & .29 & .55 \tabularnewline
\hline
Model B &        &	  &  &  & \tabularnewline
\hline
$p$$^a$& Burst             & 0.002 & 0.04& 0.100 & 0.0005 \tabularnewline
& Continuous             &   1.5e-5 & 0.0002 &  0.006&  0.06 \tabularnewline
$f > 0.97$$^b$ & Burst     & .45 & .35 & .22 & .07 \tabularnewline
& Continuous             &  .50 & .47 &  .38 & .19 \tabularnewline
DR$^c$ & Burst        & .17 & .29 & .45 & .70 \tabularnewline
& Continuous             & .08  & .16 & .26 & .55 \tabularnewline
\hline
\multicolumn{6}{l}{$^a$ Two-dimensional $(a,e)$ Kolmogorov-Smirnov $p$-values.}\tabularnewline
\multicolumn{6}{l}{$^b$ Fraction of S-stars with $e > 0.97$ at end of simulation.}\tabularnewline
\multicolumn{6}{l}{$^c$ Disruption rate (fraction) of S-stars.}\tabularnewline
\multicolumn{6}{l}{$^d$ Stars initialized at $t = 0$.}\tabularnewline
\multicolumn{6}{l}{$^e$ Stars randomly initialized between $[0,t_{\rm max}]$.}\tabularnewline
\end{tabular}
\end{center}
\label{t:ks}
\end{table}

\subsubsection{Results}

\begin{figure}
\begin{center}
			\includegraphics[height=100 mm,angle=270,trim=0.5cm 3.5cm 0cm 0cm,clip ]{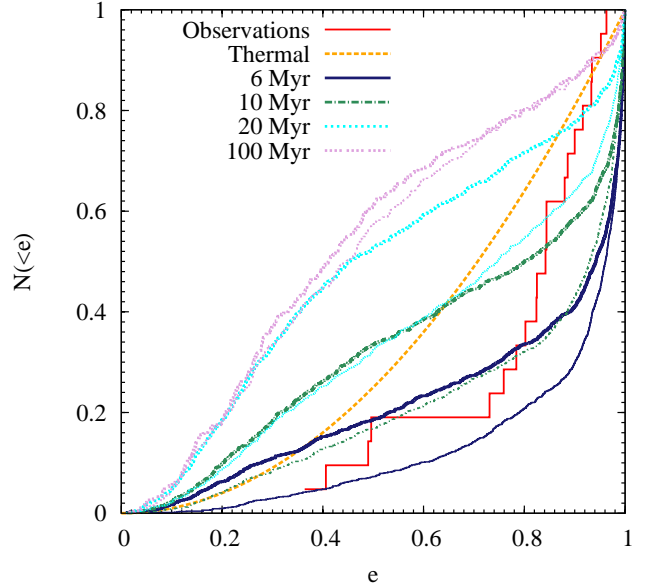}
		\caption{Cumulative distribution of eccentricities for the {\it burst} scenario in which the stars' initial time is  zero. The red step function is the observed cumulative eccentricity distribution of S-stars in the GC as found by \citet{Gil09}. The dashed orange line is a thermal distribution $N(<e)=e^2$. The other lines are results from MC simulations, where the evolution time $t_{\rm max}$ and eccentricity $e$ are indicated in the legend. Thick (thin) lines are the results for Model A (B). \label{f:Sstars_burst}}
				\end{center}
\end{figure}

\begin{figure}
\begin{center}
			\includegraphics[height=100 mm,angle=270,trim=0.5cm 3.5cm 0cm 0cm,clip]{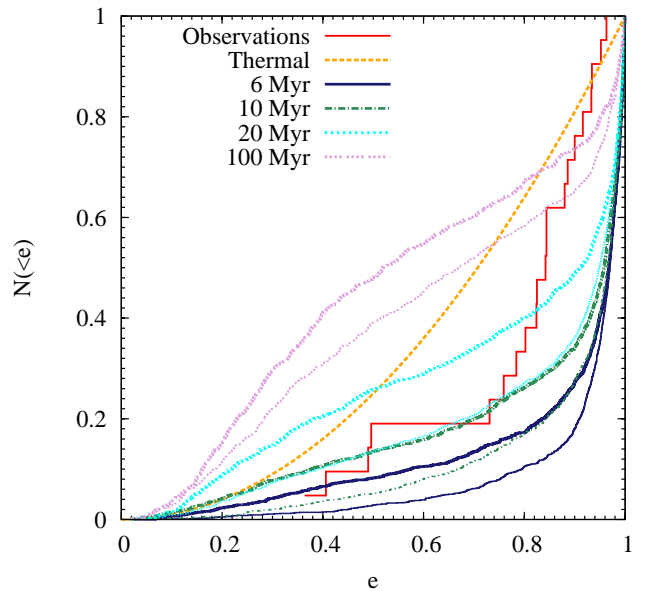}
		\caption{Cumulative distribution of eccentricities for the {\it continuous} scenario in which the stars' initial time $t$ is randomly distributed between $[0,t_{\rm max}]$. The red step function is the observed cumulative eccentricity distribution of S-stars in the GC from \citet{Gil09}. The dashed orange line is a thermal distribution $N(<e)=e^2$. The other lines are results from MC simulations, where the maximum evolution time $t_{\rm max}$ and initial eccentricity $e$ are indicated in the legend. Thick (thin) lines are the results for Model A (B). \label{f:Sstars_cont}}
		\end{center}
\end{figure}

For both models we find the cumulative eccentricity distribution of the stars at different times and compare them to the observed distribution from \citet{Gil09}; see Figures \ref{f:Sstars_burst} and \ref{f:Sstars_cont}. In Table \ref{t:ks}, we present results from nonparametric two-dimensional Kolmogorov-Smirnov (2DKS) testing between the observed cumulative distribution and the simulations. The two-sample KS test checks whether two data samples come from different distributions (low $p$ values). We also give the fraction of stars that have an eccentricity $e > 0.97$ at the end of the simulation. The total disruption rate (DR) gives an indication of how many progenitors would be needed to make up the population of S-stars observed today. 

{\it Model A}: The best fit to the S-star observations are for the {\it burst} scenario with $t_{\rm max} = 6-10 \Myr$. Greater values of $t_{\rm max}$ produce too many low eccentricity stars in our simulations; as the RR time is longer at these eccentricities stars will tend to remain at these low values. In the {\it continuous} scenario the simulations conclude with many more high eccentricity stars as they do not have as much time to move away from their original eccentricities. Hence the best-fit result for this scenario is $t_{\rm max} = 10-20 \Myr$. In both cases the 2DKS $p$-values $\ll1$, suggesting that no simulated distribution matches very well the observed. Both scenarios predict a large number ($7 - 45 \%$) of high eccentricity stars with $e > 0.97$. 

{\it Model B}: The best match to the observations is the {\it burst} scenario with $t_{\rm max} = 20 \Myr$, though the {\it continuous} scenario has a relatively high $p$-value at a time of $t_{\rm max} = 100 \Myr$. In both cases a longer $t_{\rm max}$ is preferred as this model has less mass, and hence GR precession will dominate the precession rate of these stars down to lower eccentricities, rendering RR less effective at high eccentricities. Again the $p$-values are low in most cases due to the excess of high eccentricity orbits relative to the observations. \\

For both models, simulations with $t_{\rm max}  > 100 \Myr$ produce too many low eccentricity stars with respect to the observations. Simulations with $t_{\rm max} = 200 - 400 \Myr$ have increasingly smaller $p$-values for this reason. The small number of observed low eccentricity S-stars could in principle constrain the time S-stars have spent at these radii (their ``dynamical lifetimes''). Although the S-stars can in theory have ages up to $\sim 400 \Myr$, we find that, in the binary disruption context, their dynamical lifetimes at these radii are $< 100 \Myr$. Furthermore, as noted by \citet{Per09b}, the lack of observed low eccentricity S-stars is difficult to reconcile with formation scenarios which bring the stars to these radii on initially low eccentricity orbits. The RR timescale $t_{\rm rr}$ is long and stars will retain their initial values for long times. 

\begin{figure}
\begin{center}
			\includegraphics[height=85 mm,angle=270 ]{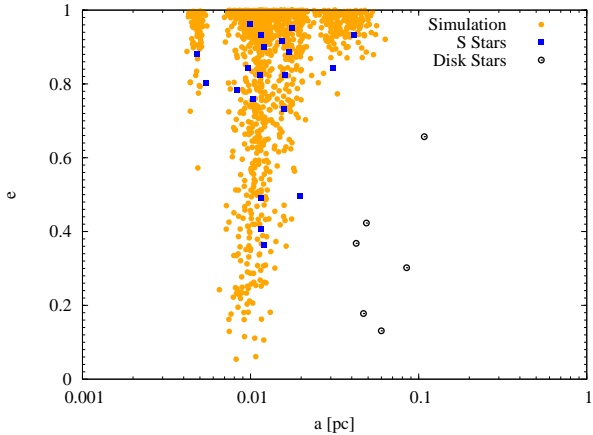}
		\caption{Scatter plot of eccentricity, $e$, of the S-stars, as a function of semi-major axis, $a$, on a log scale for the Model B {\it burst} scenario. The small orange dots represent the results of our simulations after $6 \Myr$, the blue squares are data from \citet{Gil09} and the black open circles are those S-stars consistent with being members of the CW disk.\label{f:Scatter_Sstars}}
		\end{center}
\end{figure}

For all values of $t_{\rm max}$ there is a clear discrepancy between the lack of high eccentricity S-stars orbits observed and the results from both formation scenarios and models. To further illustrate this problem we show a scatter plot from our simulations in Figure \ref{f:Scatter_Sstars}, of eccentricity $e$ as a function of semi-major axis $a$ after $6 \Myr$. We over-plot the observed S-star parameters, indicating separately those associated with the disk. It is clear that while the simulations can reproduce the range of $(a,e)$ in which the S-stars are observed, our theory over-predicts the amount of high eccentricity orbits with respect to the observations. Such orbits at these radii are expected; stars with high orbital eccentricities, although they experience large torques (Equation (\ref{e:tau})), have short coherence times (Equation (\ref{e:tphi})) and hence long RR times (Equation (\ref{e:tRR})). Consequently, their angular momentum evolution is sluggish. At any given time, we would expect to see a population of stars in this region. Therefore, we find it difficult to explain why so few high eccentricity S-stars have been observed; the maximum is $e = 0.963 \pm 0.006$ \citep[S14;][]{Gil09}; $e = 0.974 \pm 0.016$ \citep[S0-16;][]{Ghe05}.

An interesting solution to this problem is that there {\it are} S-stars with higher orbital eccentricities but that their orbital elements have not yet been determined due to the observational detection limits for stellar accelerations. \citet{Sch03} show that there is a bias against detecting high eccentricity orbits at the average S-star radius of $0.01 \pc$ (see their Figure 14). Future observations will be able to reveal if this can account for the discrepancy \citep{Wei05}. The solution on the other hand may involve a new stellar dynamical mechanism which is not accounted for in our ARMA model. \citet{Mer11} have recently discovered a new dynamical barrier in angular momentum space, termed the ``Schwarzschild barrier'', which can reflect stars orbiting close to the MBH from very high eccentricity orbits ($e > 0.99$) to less eccentric ones. This could well explain why not so many S-stars have been observed on highly eccentric orbits as RR alone predicts.

\section{Summary} \label{s:disc}

We have used several techniques to study the angular momentum evolution of stars orbiting MBHs, with a focus on secular effects. First, we carried out $N$-body simulations to generate time series of angular momentum changes for stars with various initial eccentricities. We quantified the evolution in an ARMA(1, 1) model. We then used this model to generate new, much-longer time series in our MC code, which enabled us to study the steady-state distribution of stellar orbits around an MBH, and the evolution of the angular momentum distribution of possible progenitors of the S-stars. 

We have shown that the steady-state angular momentum distribution of stars around an MBH is not isotropic when RR is a dominant relaxation mechanism. For the GC this corresponds to the innermost couple of tenths of parsecs. We note that \citet{Sch09} find a slight degree of anisotropy in the velocity distribution of late-type stars in the GC within the innermost $6$"  ($\sim0.2 \pc$). It is of particular interest to note that close to the MBH, we find that the angular momentum distribution is steeper than linear, which may have
consequences for the stability of the system. \citet{Tre05} and \citet{Pol07} showed
that if (1) the distribution rises faster than linearly, and (2) the
distribution of angular momenta vanishes at $J=0$, the system is only
neutrally stable (or even unstable if it is flattened). The second
condition is also satisfied, because of the presence of the loss cone:
$N=0$ for $J<J_{lc}$. In a later paper, \citet{Pol08} found that a non-monotonic distribution of angular momenta is required for the system to develop a ``gravitational loss cone instability''. This condition is also satisfied for a range of energies. An assessment of the importance and consequences of these instabilities is beyond the scope of this work. 

We have furthermore demonstrated the presence of a depression in the distribution of stars near MBHs due to resonant tidal disruption. This depression forms on a timescale shorter than the Hubble time, even though a steady-state solution may not yet  be reached. While it is tempting to invoke this mechanism to explain the ``hole'' in late-type stars in the GC, we stress that our fiducial galactic nucleus model, a single-mass BW distribution of stars around an MBH, is greatly simplified compared to reality and produces an outer hole radius which is too small to explain the observations. We expect the GC to be far more complex, with ongoing star formation, multi-mass stellar populations, mass segregation, physical collisions between stars and additional sources of relaxation (e.g., the circumnuclear ring, the CW disk). Nevertheless, there is merit to our approach: even with our simplified picture, we illustrate the importance of RR on stellar dynamics close to an MBH. 

Finally we have shown that the $(a,e)$ distribution of the S-stars does not match that expected by the binary disruption scenario in which the stars start their lives on highly eccentric orbits and evolve under the RR mechanism. We surmise that there are a number of high eccentricity S-stars whose orbital parameters have not yet been derived, though the ``Schwarzschild barrier'' recently discovered by \citet{Mer11} may provided a dynamical solution to this problem. We have confirmed the result by \citet{Per09b} in that it is unlikely that the S-stars were formed on low eccentricity orbits. We place a constraint on the dynamical age of the S-stars of $100 \Myr$ from the small number of low eccentricity orbits observed. 

\acknowledgements{
A.M. is supported by a TopTalent fellowship from the Netherlands Organisation for Scientific Research (NWO), C.H. is supported by a Veni fellowship from NWO, and Y.L. by a VIDI fellowship from NWO. We are very grateful to Atakan G\"urkan for his significant contribution to this project. A.M. thanks Andrei Beloborodov, Pau Amaro-Seoane and Eugene Vasiliev for stimulating discussions. We thank Tal Alexander and Ben Bar-Or for their comments on the first draft of this paper, particularly with respect to energy diffusion. Finally, we thank the anonymous referee for a comprehensive reading of this paper and insightful comments.}

\appendix

\section[]{Description of $N$-body code}\label{s:app:code}

We have developed a new integrator, specifically designed for stellar dynamical simulations in near-Keplerian potentials. A distinctive feature of this code is the separation between ``test'' particles and ``field'' particles \citep[see, e.g.,][]{Rau96}. The field particles are not true $N$-body particles; they move on Keplerian orbits which precess according to an analytic prescription (see Equation (\ref{e:app:t_cl})), and do not directly react to the gravitational potential from their neighbors. Conversely, the test particles are true $N$-body particles, moving on Keplerian orbits and precessing due to the gravitational potential of all the other particles in the system. Both test and field particles' orbits precess due to GR effects. This reduces the force calculation from $O(N^2) \rightarrow O(nN)$, where $N$ is the number of field particles and $n$ is the number of test particles in the simulation. 

The main difference between our code and many others, is that instead of considering perturbations on a star moving on a straight line with constant velocity, we consider perturbations on a star moving on a precessing Kepler ellipse. Our algorithm is based on a mixed-variable symplectic method \citep{WH91, Kin91, Sah92}, so called due to the switching of the coordinate system from Cartesian (in which the perturbations to the stellar orbit are calculated) to one based on Kepler elements (to determine the influence of central Keplerian force). These forms of symplectic integrator split the Hamiltonian, $H$, of the system into individually integrable parts, namely a dominant Keplerian part due to the central object, i.e., MBH, and a smaller perturbation part from the surrounding particles, i.e., stellar cluster,
\begin{equation}
H = H_{\rm Kepler} + H_{\rm interaction}.
\end{equation}
The algorithm incorporates the well-known leapfrog scheme where a single time step ($\tau$) is composed of three components, kick ($\tau /2$)-drift ($\tau$)-kick ($\tau /2$). The {\it kick} stage corresponds to $H_{\rm interaction}$ and produces a change of momenta in fixed Cartesian coordinates. The {\it drift} stage describes motion under $H_{\rm Kepler}$ and produces movement along unperturbed Kepler ellipses. To integrate along these ellipses, we exploit the heavily-studied solutions to Kepler's equation. This is a less computationally efficient method than direct integration, but accurate to machine precision. This is of utmost importance in simulations involving RR as spurious precession due to build-up in orbital integration errors will lead to erroneous results. The computation is performed in Kepler elements, using Gauss' $f$ and $g$ functions. We use universal variables which allow the particle to move smoothly between elliptical, hyperbolic, and parabolic orbits, making use of Stumpff functions. For details we refer the reader to \citet{Danby92}. For the simulations presented in this paper, we use a fourth-order integrator \citep{Yosh90} which is achieved by concatenating second order leapfrog steps in the ratio $(2 - 2^{1/3})^{-1}\!:\!-\!2^{1/3}(2 - 2^{1/3})^{-1}\!:\!(2 - 2^{1/3})^{-1}$. 

\begin{figure}
\begin{center}
			\includegraphics[height=85 mm,angle=270 ]{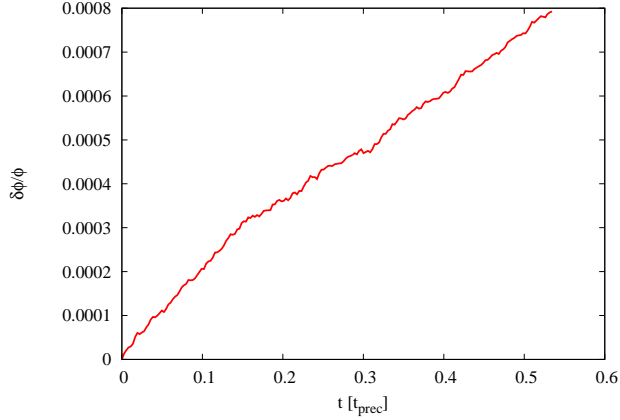}
		\caption{Fractional difference between the theoretical value for apsidal precession and that achieved by our integrator for a test star of eccentricity $e = 0.99$ as a function of time in units of the precession time of the test-star. The angle through which the test star has precessed is given by $\phi$. The analytical formula is adjusted for the test star's wandering in semi-major axis.
		\label{f:prec_rate_0p99}}
		\end{center}
\end{figure}

\begin{figure}
\begin{center}
			\includegraphics[height=85 mm,angle=270 ]{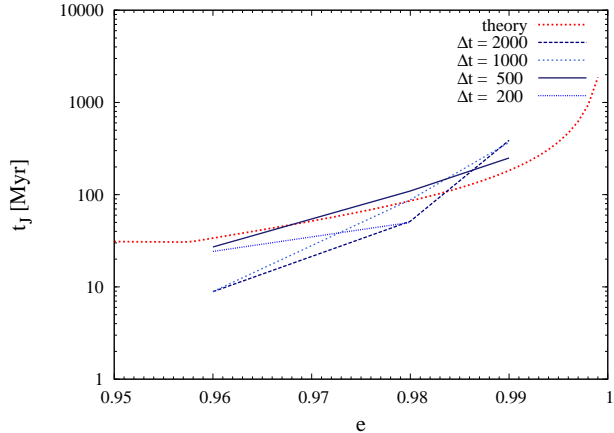}
		\caption{Timescale for angular momentum $J$ changes of the order of the circular angular momentum for massless test stars in our $N$-body simulations, presented on a log scale of time in units of Myr; see Equation (\ref{e:diffusion_J}). We take several $\Delta t$ values to get order-of-magnitude estimates for this timescale. Over-plotted in red is the theoretical expectation for this timescale based on Equation (\ref{e:tRR}).
		\label{f:0p96-0p99}}
		\end{center}
\end{figure}

We use adaptive time stepping to resolve high eccentricity orbits at periapsis, and to accurately treat close encounters between particles, hence our integrator is not symplectic. To reduce the energy errors incurred with the loss of symplecticity, we implement time symmetry in the algorithm. These steps must be sufficient to avoid artificial apsidal precession of stellar orbits, particularly at high eccentricity where Wisdom-Holman integrators are known to perform poorly \citep{Rau99}. To demonstrate this we compare the rate of change, as a function of time, of the direction of the eccentricity vector of a test star using our $N$-body integrator with our analytical prediction. Figure \ref{f:prec_rate_0p99} shows our results in terms of the angle through which the test star has precessed, $\phi$, for $e = 0.99, a  = 0.01 \pc$ orbits in the gravitational potential as described in Section \ref{ss:Model_GC}. We take the average of nine simulations for half a precession time of the test star. Extrapolating to a full precession time we find a fractional difference of $\delta \phi / \phi = 0.00148$ between our analytical estimate and that of the test star precession rate. This verifies that the long RR timescale at high eccentricities, as shown in Figure \ref{f:trr_tnr}, is a result of attested rapid apsidal precession. To further demonstrate the accuracy of the integrator with respect to the apsidal precession rate, we run additional calibration simulations for high-eccentricity orbits ($e = 0.96, 0.98$) in order to compare their angular momentum relaxation timescale (calculated using Equation (\ref{e:diffusion_E})) with the formula we derive for RR in Equation (\ref{e:tRR}). Figure \ref{f:0p96-0p99} demonstrates the agreement between increase of the RR timescale with eccentricity as retrieved from the simulations and that expected from theory. \\

An additional feature of the code is the option to implement ``reflective'' walls, which bounce field particles back into the sphere of integration. Without these walls, particles on eccentric orbits nearing apoapsis move outside the sphere of interest and alter the density profile of the background cluster of particles, and hence the (stellar) potential. With the walls ``on'', a particle reaching the reflective boundary is reflected directly back but with a different random initial location on the sphere, to avoid generating fixed orbital arcs in the system which would artificially enhance RR. The reflective walls are most useful in situations where we want to examine how a test particle reacts in a defined environment with a specific background (stellar) potential. The prescription we use for the time step ($\tau$) is based on orbital phase, the collisional timescale between particles and free-fall timescale between particles. We treat gravitational softening between particles with the compact $K_2$ kernel \citep{Deh01}. 

We have parallelized our code for shared-memory architecture on single multiple-core processors using OPENMP.

\section[]{Energy evolution and cusp formation}\label{s:app:energy}

Many different definitions of relaxation times are used in the literature. In Section \ref{ss:Sim_Results} we consider the energy diffusion timescale $t_E$, defined as $t_E\equiv E^2/D_{\rm EE}$ where $D_{\rm EE} \equiv \langle (\Delta E)^2 \rangle / \Delta t$ is the energy diffusion coefficient. This is the timescale for order unity steps in energy. In order to be explicit, we now compare our timescale (see Figure \ref{f:trr_tnr} and Equation (\ref{e:diffusion_E})) with a commonly used reference time for relaxation processes defined in \citet{Spi87}:
\begin{equation}
\begin{split}
t_r &= \frac{1}{3} \frac{v_M^2}{\langle(\Delta v_{||})^2 \rangle_{v=v_M}} \\
&= 0.065\frac{v_M^3}{(G m)^2 n \ln{\Lambda}}.
\end{split}
\end{equation}
We note that this is equivalent to the timescale  given by \citet{Bin08} for a Maxwellian velocity distribution, $v_M = \sqrt{3}\sigma$, where $v_M (r)$ is the mean velocity of a star and $\sigma (r)$ is the one-dimensional velocity dispersion of the stellar distribution at radius $r$,
\begin{equation} \label{e:tr}
t_r = \frac{0.34 \sigma^3}{(Gm)^2 n \ln{\Lambda}}.
\end{equation}
The velocity dispersion within a stellar cusp with power-law density profile $n(r) \propto r^{-\alpha}$ can be calculated from the Jeans equation for spherical isotropic systems as
\begin{equation}
\sigma^2(r) \equiv - \frac{G}{n(r)} \int_0^r \frac{dr' M(r') n(r')}{r'^2} = \frac{1}{1+\alpha} v_c^2
\end{equation}
where $M(r)$ is the mass within a radius $r$ and $v_c$ is the circular velocity. For the comparison of $t_E$ and $t_r$ we take Keplerian parameters for the circular velocity at a given radius such that
\begin{equation}
\sigma^2 = {1\over1+\alpha}{G\Mbh\over r}
\end{equation}
\citep{Ale03c}, and set the Coulomb logarithm to $\Lambda=\Mbh/m$ \citep{Bah76}. For the cusp model used in our $N$-body simulations we find from Equations (\ref{e:tNR}) and (\ref{e:tr})
\begin{equation}\label{e:tE}
t_E \equiv {E^2\over D_{\rm EE}} = 0.26\left({\Mbh\over m}\right)^2{1\over N_<}{1\over \ln\Lambda}P = 2.18 ~ t_r, 
\end{equation}
with $t_r = 156\,{\rm Myr}$, $t_E=340\,{\rm Myr}$ for $r = 0.01 \pc$. Thus with this definition, the energy of a star changes by order unity in roughly twice a relaxation time. \\

We compare our result for the energy relaxation time to that found by \citet{Eil09} using an entirely different integrator. Their $N$-body simulations incorporate a fifth-order Runge-Kutta algorithm with optional pairwise KS regularization \citep{Kus65}, without the need for gravitational softening. We use their numerically derived value (see their Table 1) of $\langle \alpha \rangle \equiv \langle \alpha_\Lambda/  \sqrt{\ln \Lambda} \rangle$, averaged over all stars in the cluster, and their expression for the energy relaxation time $T^E_{\rm NR} = (\Mbh/m)^2 P/(N \alpha_{\Lambda}^2)$, and compare with our Equation (\ref{e:tNR}) with its own numerically derived parameter.  We find an energy relaxation timescale which is a factor of $\sim$ four longer than in their paper (i.e., their energy diffusion timescale is about half of $t_r$). Our maximum radius at $0.03 \pc$ means we do not simulate several ``octaves'' in energy space which could contribute to energy relaxation. In a Keplerian cusp, the variance in the velocity changes of a star, $(\Delta v)^2$, is not independent of radius $r$ but scales with the power-law density index $\alpha$. In our simulations $\alpha = 7/4$, and hence the number of octaves from the cutoff radius, $r = 0.03 \pc$, out to the radius of influence, $r_h = 2.31 \pc$, can account for at most a factor of $\sim$ two difference in $t_{\rm E}$. The inner cutoff radius at $r = 0.003 \pc$ contributes a smaller factor of difference to the energy diffusion timescale.\\

\begin{figure}
\begin{center}
			\includegraphics[height=85 mm,angle=270]{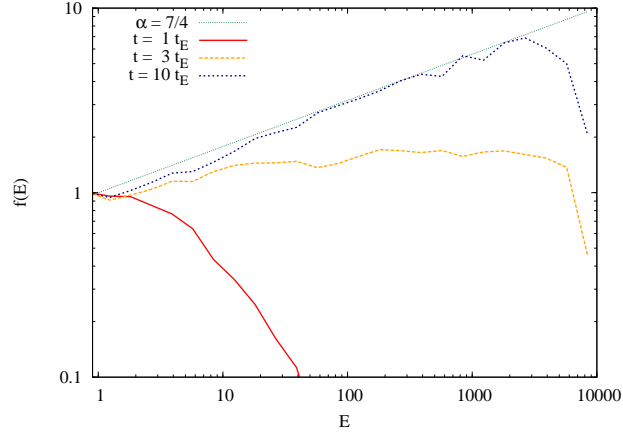}
		\caption{Stellar distribution function $f(E)$ for single mass stars as a function of energy $E$ for different times $ t = (1 - 10) ~ t_E$, where $t_E$ is the Keplerian energy diffusion timescale evaluated at $E = 1$. steady-state is reached within $10 ~t_E$, the solution of which is a BW cusp with $\alpha = 7/4$, represented by the dotted green line. 
		\label{f:cusp}}
				\end{center}
\end{figure}

We now consider the timescale for building a cusp due to energy diffusion. For this purpose, we use our MC code without angular momentum evolution. In Figure \ref{f:cusp}, we show the distribution function for several times, where the system was initialized such that all stars start at $E=1$, corresponding to half the radius of influence. {\it From this figure, we find that it takes about $10 ~t_E (E = 1, r  = 0.5 ~ r_h)$ or 8 $t_E (r = r_h)$  to form a cusp.} The result that it takes much longer than $t_E$ is also implicitly found in \citet{Bah76}. They give an expression for $D_{\rm EE}$ (called $c_2(E, t)$ in their paper); evaluating the pre-factor in their expression due to only the contribution of bound stars in steady-state, one finds that $c_2(E=1)=16/t_{\rm cusp}^{\rm BW}$, where $t_{\rm cusp}^{\rm BW}$ is the reference time in which \citet{Bah76} find a cusp is formed. \\

\citet{Bau04a} and \citet{Pre04} were the first to show the formation of a cusp in direct $N$-body simulations. We use Equations (\ref{e:tr}) and (\ref{e:tE}) to calculate the cusp formation times in Table 1 of \citet{Pre04} in terms of our energy diffusion time. We find that in their simulations a cusp forms on a timescale of the order of $10 ~t_E$. We conclude that energy diffusion proceeds in their simulations at a rate similar to ours. However, \citet{Pre04} state that a cusp forms within one relaxation time. These two statements, apparently contradictory, are reconciled in two ways. First, \citet{Pre04} define the relaxation time at a radius such that the mass in stars is equal to twice that of the MBH. Second, they include the non-Keplerian potential in their calculation of the relaxation time. It is valid to assume Keplerian parameters for the stellar velocity dispersion at the distances we are interested in close to the MBH. However, neglecting the contribution of the surrounding stellar mass leads to a significant numerical difference in the calculation of the relaxation time near the radius of influence compared with those which incorporate the non-Keplerian potential. \\

To verify this, and to reconcile our conclusions, we adapt our definition of the energy diffusion time by incorporating the non-Keplerian potential,
\begin{equation}\label{eq:v3'}
v_{c}^{3} = \left( \frac{G[\Mbh + M(<r)]}{r} \right)^{3/2} = \left(\frac{G\Mbh}{r}\right)^{3/2} \left[ 1 + \frac{M_h}{\Mbh r_h^{3- \alpha}} r^{3-\alpha}\right]^{3/2},
\end{equation}
\begin{equation}\label{eq:tE'}
t_{\rm E}' = \left(1 + A E^{\alpha - 3}\right)^{3/2} t_E, 
\end{equation}
where $A = 2^{\alpha - 3} M_h/\Mbh$, $M_h$ is the mass inside the radius of influence $r_h$, $t_{\rm E}'$ is the new energy diffusion timescale and we have used the relation $E = r_h/2a$. The new energy diffusion coefficient is
\begin{equation}\label{eq:dEE'}
D_{\rm EE}'  = \left(1 +  A E^{\alpha - 3}\right)^{-3/2} D_{\rm EE}.
\end{equation}

\begin{figure}
\begin{center}
			\includegraphics[height=85 mm,angle=270]{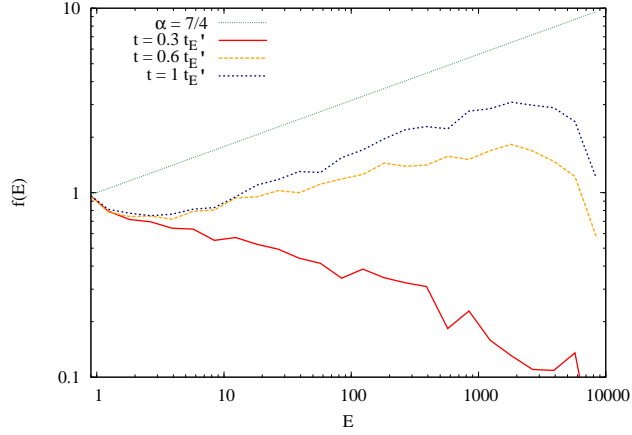}
		\caption{Stellar distribution function $f(E)$ for single mass stars as a function of energy $E$ for different times $ t = (0.3 - 1) ~ t_E'$, where $t_E'$ is the energy diffusion timescale which incorporates the non-Keplerian stellar potential and is evaluated at $E = 0.287, r  = 4.023 \pc$. The stars start in a delta function at $E = 1$ (and hence generate an artificial peak at this position) and form a cusp within one relaxation time $1 ~t_E'$.
				\label{f:cusp_nonkep}}
				\end{center}
\end{figure}

In Figure \ref{f:cusp_nonkep} we show the results of MC simulations in which the new energy diffusion coefficient is used. The system was again initialized such that all stars start at $E=1$. A BW cusp ($\alpha = 7/4$) forms in about one relaxation time evaluated at $E = 0.287$, the energy at which $M_h = 2 \Mbh$. This result emphasizes the importance of taking the stellar potential into account for the relaxation timescale. \\

For a simplistic single mass ($m = 1\,\Mo$) GC model (see Section \ref{ss:Model_GC}), we find that at the radius of influence the Keplerian energy diffusion timescale is $t_E = 12\Gyr$ and a cusp forms in $\sim100\Gyr$. Alternatively we find from Equation (\ref{eq:tE'}) a cusp formation time of $t_E' \sim 90 \Gyr$. We stress that the time for cusp formation is much longer than the Hubble time, even if we halve it to account for the outer cutoff in radius in our $N$-body simulations which suppresses the energy diffusion timescale. This is in contradiction to what is often claimed; one cause of the discrepancy is that in, for example, \citet{Bah76} and \citet{Lig77}, energy diffusion proceeds at a rate exceeding the relaxation rate by a factor $>10$, in contrast to what we find in our $N$-body simulations. We note here that if energy diffusion did in fact proceed at this higher rate, then the energy diffusion timescale for the S-stars in the GC would be on the order of $\sim 20 \Myr$ which would have profound consequences for the $(E, J)$ evolution of the cluster. It thus appears that the GC cannot have reached a steady-state \citep[see also][]{Mer10}, even though mass-segregation enhances the rate of cusp formation by a factor of a few \citep{Pre10, Hop10}.

\section[]{Precession due to power-law stellar cusp}\label{s:app:prec}

The presence of a nuclear stellar cluster in the vicinity of an MBH adds a small correction $\delta U(r)$ to the Keplerian gravitational potential energy $U = -\alpha/r$ of the system. As a consequence, paths of finite motion (i.e., orbits) are no longer closed, and with each revolution the periapsis of a star is displaced through a small angle $\delta \phi$. Given a power-law density profile of the stellar cluster $n(r) \propto r^{-\alpha}$, we can calculate this angle using the formula
\begin{equation} \label{e:app:dU}
\delta \phi = \frac{\partial}{\partial L} \left( \frac{2}{L} \int^{\pi}_0 r^2 \delta U d\phi \right),
\end{equation}
where $L = \sqrt{G \Mbh a ~(1 - e^2)}$ is the unnormalized angular momentum of a Keplerian orbit  \citep{Lan69}. We continue to use specific parameters in the following equations. The stellar mass within radius $r$ is
\begin{equation} \label{e:m(r)}
M(r) = M_0 \left({r \over r_0} \right)^{3 - \alpha} ,
\end{equation}
$M_0$, $r_0$ being proportionality constants which we normalize to the radius of influence, $r_0 = r_h$, $M_0 = \Mbh$. Thus the potential felt by a massless particle due to this enclosed mass is
\begin{equation} \label{e:app:dU_equals}
\begin{split}
\delta U &= - \int_0^r F(r') \partial r'\\ 
&= \frac{G \Mbh r_h^{\alpha - 3}}{(2 - \alpha)} r^{2 - \alpha} \quad (\alpha \ne 2).
\end{split}
\end{equation}
Inserting $\delta U$ into Equation (\ref{e:app:dU}) yields
\begin{equation}\label{e:app:dphi_prec}
\delta \phi = \frac{G \Mbh r_h^{\alpha - 3}}{(2 - \alpha)} \frac{\partial}{\partial L} \left( \frac{2}{L} \int^{\pi}_0 r^{4 - \alpha} d\phi \right).
\end{equation}
We apply a change of coordinates here, replacing $L$ with normalized angular momentum $J = \sqrt{1 - e^2}$ , such that $L = \sqrt{G \Mbh a} J$.
Expressing radius $r$ in terms of Kepler elements, semi-major axis $a$, eccentricity $e$ and angle from position of periapsis $\phi$, 
\begin{equation}
r = a \frac{1-e^2}{1 + e \cos{\phi}} = a \frac{J^2}{1 + \sqrt{1 - J^2} \cos{\phi}},
\end{equation}
yields
\begin{equation}
\begin{split}
\delta \phi &= {2 \over (2 - \alpha)} \left( {a \over r_h} \right)^{3 - \alpha} f(e, \alpha)  \\
&= \frac{2}{(2 - \alpha)}   f(e, \alpha) \left[ {N_< m \over \Mbh} \right],
\end{split}
\end{equation}
where $m$ is the mass of a single star, $N_< $ is the number enclosed within radius $r$ and
\begin{equation}
f(e, \alpha) = \frac{\partial}{\partial J} \left( \frac{1}{J} \int^{\pi}_0 \left[ \frac{J^2}{1 + \sqrt{1 - J^2} \cos{\phi}} \right]^{4 - \alpha} d\phi \right).
\end{equation}
The function $f(e, \alpha)$ is calculated numerically, and fit, for a given $\alpha$. This returns a cluster precession time of 
\begin{equation}\label{e:app:t_cl}
t^{\rm cl}_{\rm prec} = \pi (2 - \alpha) f(e, \alpha)^{-1} \left[ {\Mbh\over N_< m}P(a) \right]\quad (\alpha \ne 2),
\end{equation}
where $P(a)$ is the period of an orbit with semi-major axis $a$.

\section[]{Equations of ARMA(1,1) model}\label{s:app:eqns}

We will use Equations (\ref{e:DJ1}) and (\ref{e:iid}) in our calculations; for clarity we repeat them here, dropping the label ``1'',
\begin{equation}\label{e:app:DJ}
\Delta J_t = \phi \Delta J_{t-1} + \theta \epsilon_{t-1} + \epsilon_t,
\end{equation}
\begin{equation}\label{e:app:iid}
\langle\epsilon\rangle=0;\quad \langle\epsilon_t\epsilon_s\rangle=\sigma^2\delta_{t, s}.
\end{equation}

\subsection{Variance}

\begin{equation}\label{e:app:DJ2}
\begin{split}
\langle \Delta J_t^2\rangle = \phi^2 \langle \Delta J_{t-1}^2\rangle &+ 2 \phi \theta \langle \Delta J_{t-1} \epsilon_{t-1} \rangle + 2 \phi \langle \Delta J_{t-1} \epsilon_t \rangle \\
&+ \theta^2 \langle \epsilon_{t-1}^2 \rangle + 2 \theta \langle \epsilon_{t-1}\epsilon_{t}\rangle + \langle \epsilon_{t}^2 \rangle
\end{split}
\end{equation}
Using $\langle \Delta J_t^2\rangle = \langle \Delta J_{t-1}^2\rangle$, expanding $\Delta J_{t-1}$ in the second and third terms, and applying $\langle\epsilon_t\epsilon_s\rangle=\sigma^2\delta_{t, s}$ yields
\begin{equation}
\langle \Delta J_t^2\rangle (1-\phi^2) = \sigma^2 (2 \phi \theta + \theta^2 + 1) ,
\end{equation}
which returns Equation (\ref{e:sigvar}):
\begin{equation}\label{e:app:sigvar}
\langle \Delta J_t^2\rangle = {1 + \theta^2 + 2\theta\phi\over 1-\phi^2}\sigma^2.
\end{equation}

\subsection{Autocorrelation function}

The ACF for the ARMA model as described by Equation (\ref{e:app:DJ}), is defined as
\begin{equation}\label{e:app:acf_def}
\rho_t = { \langle \Delta J_{s + t} \Delta J_s \rangle \over \langle \Delta J_t^2\rangle} \quad (t > 0) .
\end{equation}
Expanding the numerator gives
\begin{equation}\label{e:app:acf1}
\begin{split}
\langle \Delta J_{s + t} \Delta J_s \rangle &= \phi^2 \langle \Delta J_{s + t -1} \Delta J_{s-1} \rangle + \phi \theta \langle \Delta J_{s + t -1} \epsilon_{s-1} \rangle \\ & + \phi \langle \Delta J_{s + t -1} \epsilon_{s} \rangle +  \phi \theta \langle \Delta J_{s -1} \epsilon_{s+t-1} \rangle \\ &+ \phi \langle \Delta J_{s-1} \epsilon_{s+t} \rangle + \theta^2 \langle \epsilon_{s+t-1} \epsilon_{s-1} \rangle \\& + \theta \langle \epsilon_{s+t-1} \epsilon_{s} \rangle + \theta  \langle \epsilon_{s+t} \epsilon_{s-1} \rangle +  \langle \epsilon_{s} \epsilon_{s+t} \rangle,
\end{split}
\end{equation}
where $\langle \Delta J_{s + t} \Delta J_s \rangle = \phi^2 \langle \Delta J_{s + t -1} \Delta J_{s-1} \rangle$. Recursively expanding the $\Delta J$ terms, and again using $\langle\epsilon_t\epsilon_s\rangle=\sigma_1^2\delta_{t, s}$, reduces the fourth and fifth terms to zero, and greatly simplifies the expression to
\begin{equation}\label{e:app:acf2}
\langle \Delta J_{s + t} \Delta J_s \rangle = {\sigma^2 \phi^t (\theta + \phi) (\theta + {1/\phi}) \over 1 - \phi^2}.
\end{equation}
Normalizing by the variance returns Equation (\ref{e:acf}):
\begin{equation}\label{e:app:acf}
\rho_t = \phi^t\left[1 + {\theta/\phi\over 1 + (\phi+\theta)^2/(1-\phi^2)}\right]\quad\quad(t>0) .
\end{equation}

\subsection{Variance at coherence time $t_\phi$}

To calculate the variance at the coherence time, $\langle \Delta J_\phi^2 \rangle$, we begin by calculating the variance after some time $t$,
\begin{equation}\label{e:app:DJtACF}
\begin{split}
\left\langle\left( \sum_{n=0}^t \Delta J_n^2 \right)\right\rangle &= \int_0^t \int_0^t \langle \tau(t_1) \tau(t_2) \rangle dt_1 dt_2 \\
&=  \langle \Delta J_t^2 \rangle \int_0^t \int_0^t \rho_{(t_1 - t_2)}  dt_1 dt_2 .
\end{split}
\end{equation}

Here $ \rho_{(t_1 - t_2)}$ is the ACF of the torque at $(t_1 - t_2) > 0$, normalized with $\langle \Delta J_t^2 \rangle$ to be comparable with Equation (\ref{e:acf}),
\begin{equation}
\rho_{(t_1 - t_2)} = \phi^{(t_1 - t_2)}\left[1 + {\theta/\phi\over 1 + (\phi+\theta)^2/(1-\phi^2)}\right].
\end{equation}
From Equation (\ref{e:phith}) we can write
\begin{equation}\label{e:app:phith}
\phi^{(t_1 - t_2)}  = \exp\left[-{(t_1 - t_2) \over t_{\phi}}\right].
\end{equation}
Inserting this into Equation (\ref{e:app:DJtACF}), writing 
\begin{equation}\label{e:app:A} 
A = \langle \Delta J_t^2 \rangle \left[1 + {\theta/\phi\over 1 + (\phi+\theta)^2/(1-\phi^2)}\right] 
\end{equation}
and taking an upper limit of $t = t_{\phi}$ yields
\begin{eqnarray}\label{e:app:DJphiacf}
\langle \Delta J_\phi^2 \rangle &=& A \int_0^{t_{\phi}} dt_1 \int_0^{t_{\phi}}  \exp\left[-{(t_1 - t_2) \over t_{\phi}}\right] dt_2\nonumber\\
&=& \sigma^2 \left( {t_{\phi} \over P} \right)^2 \frac{(\theta + \phi) ( \theta + 1/\phi)}{1 - \phi^2} .
\end{eqnarray}
where we have approximated $(e + 1/e - 2) \approx 1$.

\label{lastpage}
\end{document}